\def\ARXIVVERSION{}

\documentclass[10pt]{article}

\ifdefined\ARXIVVERSION
\usepackage[preprint]{tmlr}
\else
\usepackage{tmlr}
\fi

\usepackage{amsmath,amssymb,amsfonts}
\usepackage{mathtools}
\usepackage{amsthm}
\usepackage{graphicx}
\usepackage{booktabs}
\usepackage{xcolor}
\usepackage{url}
\usepackage{hyperref}

\usepackage[normalem]{ulem}

\newcommand{\revdel}[1]{}

\newcommand{\bm}[1]{{\mbox{\boldmath $#1$}}}
\newcommand{\argmax}{\mathop{\rm arg~max}\limits}
\newcommand{\argmin}{\mathop{\rm arg~min}\limits}

\newtheoremstyle{nonitalic}%
  {3pt}{3pt}{\normalfont}{}{\bfseries}{.}{.5em}{}
\theoremstyle{nonitalic}
\newtheorem{theorem}{Theorem}
\newtheorem{lemma}{Lemma}
\newtheorem{corollary}{Corollary}

\newtheorem{definition}{Definition}
\newtheorem{remark}{Remark}
\newtheorem{assumption}{Assumption}
\newtheorem{proposition}{Proposition}

\title{The Reliability Function of Error-Correcting Output Codes\\with Noisy Binary Classifiers}

\author{%
  \name Gendo Kumoi \email kumoi@vos.nagaokaut.ac.jp \\
  \addr Department of Information and Management Systems Engineering\\
  Nagaoka University of Technology 
  \AND
  \name Hideki Yagi \email h.yagi@uec.ac.jp
  \\
  \addr Department of Computer and Network Engineering\\
  University of Electro-Communications 
  \AND
  \name Manabu Kobayashi \email mkoba@waseda.jp
  \\
  \addr Center for Data Science\\
  Waseda University
  \AND
  \name Shigeichi Hirasawa \email hira@waseda.jp
  \\
  \addr Center for Data Science\\
  Waseda University
}

\def\month{MM}
\def\year{YYYY}
\def\openreview{\url{https://openreview.net/forum?id=XXXX}}

\begin{document}

\maketitle

\begin{abstract}
Error-Correcting Output Coding (ECOC) builds a multi-class classifier from binary classifiers.
We study the excess error rate of ECOC over the Bayes (MAP) classifier when each binary classifier outputs a posterior estimate whose log-odds is corrupted by Gaussian noise.
Recasting decoding as a test over a Gaussian channel, we derive an upper bound from the union and Chernoff bounds and a lower bound from the comparison with the nearest competing category and the Mills ratio, the ECOC analogues of achievability and converse bounds in channel coding.
The bounds show that a large Hamming distance is both necessary and sufficient for reducing the excess error rate.
The excess error rate stays bounded away from zero, by a constant that depends only on the two largest posteriors and the noise level and not on the number of categories $M$, whenever the distance between the code words of the two most probable categories is bounded; for equidistant codes this distance is $d_{\min}$.
Conversely, it is at most $M-1$ times a factor that is exponentially small in the minimum distance $d_{\min}$, and it vanishes whenever $\kappa\,d_{\min}-\ln(M-1)\to\infty$ for an explicit constant $\kappa$ (for general codes, under a condition on the posteriors).
For $n$-vs-rest codes with $n\geq2$, a family of equidistant codes, the upper and lower exponents coincide asymptotically, which determines the reliability function of ECOC in closed form.
The results extend to correlated Gaussian noise with the Hamming distance replaced by an effective distance; when noise is shared at the level of categories, the effective distance stays bounded and the excess error rate does not vanish for any code word table.
In a semi-synthetic setting that satisfies the assumptions exactly, the measured decay agrees with the predicted exponent.
On ResNet-50 features of CIFAR-10/100 and ImageNet-1k, the excess error over a log-likelihood-decoding reference decreases with $d_{\min}$ and, for the larger numbers of classes, levels off at a moderate distance, while codes with $d_{\min}=1$ degrade as $M$ grows.
\end{abstract}

\section{Introduction}
A multi-class classification problem ($M\geq3$ categories) can be solved by combining several binary classifiers, a method called Error-Correcting Output Coding (ECOC) \citep{Dietterich1994-mt}.
ECOC is widely used for its performance and ease of implementation, and loss-based decoding unifies it with other reductions of multi-class to binary problems \citep{Allwein2000-mi}.
Its performance depends on which binary classifiers are combined; effective combinations have been demonstrated empirically on many datasets \citep{Dietterich1994-mt,Rifkin2004-sk,Escalera2010-pa,Rocha2014-ey}, but they depend strongly on the dataset, and no general principle governing these choices has been established.

Conditions for the optimal combination have been given under the noiseless assumption that each binary classifier estimates the true posterior \citep{james1998error,Rifkin2004-sk,kumoi2021IPSJ}.
\citet{kumoi2022SMC,kumoi2023IJNS} proposed a framework that evaluates noisy binary classifiers analytically and showed the effect of the Hamming distance on the pairwise error probability --- the error rate between two categories.
That result, however, was confined to the pairwise error probability and did not reach the classification error rate over all $M$ categories, leaving open how increasing the Hamming distance reduces the classification error rate of ECOC, and whether that contribution is both sufficient and necessary.

This paper answers these questions for the \emph{excess error rate} of ECOC, the amount by which its error rate exceeds that of the Bayes (MAP) classifier (the excess risk under the 0--1 loss; \citealp{bartlett2006convexity,tewari2007consistency}).
Our contributions are as follows.
(i)~We recast ECOC decoding under log-odds noise as a test over a Gaussian channel and reduce the excess error rate to the probability that ECOC fails to select the MAP category (Section~\ref{sec:channel}).
(ii)~We derive an upper bound (Section~\ref{sec:upper}) and a lower bound (Section~\ref{sec:lower}) that give, for every code word table, the sufficiency of a large minimum distance and the necessity of a large distance between the code words of the two most probable categories, the latter uniformly in $M$.
(iii)~For $n$vR codes ($n\geq2$), a family of equidistant codes, the two exponents coincide asymptotically, which determines the reliability function of ECOC in closed form (Section~\ref{sec:tight}).
(iv)~We extend the results to an arbitrary Gaussian noise covariance through an effective distance, under which common-mode correlation is harmless and category-level noise bounds the attainable gain (Section~\ref{subsec:correlated}).
(v)~Semi-synthetic and real-image experiments confirm the predicted exponent and the decay of the excess error rate with the minimum distance (Section~\ref{sec:experiments}), and we identify the factors that limit the quantitative agreement on real data (Section~\ref{sec:discussion}).
All lemmas, propositions, corollaries and theorems of Sections~\ref{sec:ECOC}--\ref{subsec:correlated}, with the proofs in the appendices, are machine-checked in Lean~4 with Mathlib, from the Gaussian noise model and the decoder to the reliability function (\ifdefined\ARXIVVERSION ancillary files of this arXiv submission\else supplementary material\fi); the numerical and experimental results are not part of the formalization.

\textbf{Related work.}
Regret analyses of reductions bound the excess error of a multi-class classifier by the excess error of the binary problems it is reduced to, for every distribution.
For ECOC with Hamming decoding the multi-class training error is at most $2N/d_{\min}$ times the average binary error rate \citep[Corollary~2]{Allwein2000-mi}, and error-limiting reductions bound the multi-class error by a constant multiple of the binary error \citep{beygelzimer2005error}; modified reductions instead transform binary \emph{regret} into multi-class regret, at the cost of a square root \citep{langford2005sensitive}, or with only a constant-factor increase and $O(\log M)$ computation per example \citep{beygelzimer2009tournaments}; \citet{ramaswamy2014consistency} characterize when output-code methods with surrogate losses are consistent.
In these worst-case guarantees the minimum distance enters only through a constant factor, whereas under a noise model it controls an exponential rate.
\citet{nguyen2021ensemble} showed, assuming that the binary decisions err independently, that the error rate of Hamming decoding decays exponentially in the code length.
We instead model the noise in the posterior estimates that the binary classifiers output, decode by likelihood, and bound the excess error over the MAP classifier from both sides, which identifies the exponent and its dependence on the posterior.
For codes drawn at random, \citet{kobayashi2024average} bounded the average probability that ECOC disagrees with MAP estimation by an exponent of the form $N(c-R)$ with rate $R=\ln M/N$; our results concern fixed (constructive) code word tables, and the sufficiency condition $\kappa\,d_{\min}-\ln(M-1)\to\infty$ is its counterpart.
Output codes with only $\lceil\log_2 M\rceil$ bits are also used to reduce the output layer of neural sequence models over large vocabularies, where plain binary codes have been found insufficient and redundancy from an error-correcting code improves accuracy \citep{oda2017neural}; our necessity result is consistent with this observation.

\section{Problem Setting}\label{sec:ECOC}

\subsection{ECOC and code word tables}
The classification problem is to estimate the category $c_i\ (i \in \mathcal{C} \triangleq\{1,2,\ldots,M\})$ corresponding to given input data $\bm{x} \in \mathbb{R}^D$, with $M \ge 3$.
In ECOC the category is estimated from the outputs of binary classifiers constructed according to a predefined code word table \citep{Dietterich1994-mt,Rifkin2004-sk,Escalera2010-pa}.
A code word table is an $M \times N$ matrix $W$ with entries $w_{ij} \in \{0,1\}$ in which each row is the code word of a category and each column defines a binary classifier.
We denote by $f_j$ the binary classifier constructed from the $j$-th column of $W$ and let $\mathcal{F}\triangleq\{1,2,\ldots,N\}$ be the set of binary classifiers.

We assume the following properties of $W$.
The first two entail no loss of generality, since a constant column defines no classifier and complementing a column only relabels the two sides of a classifier; the last two are requirements.
\begin{itemize}
    \item No column has all entries equal to 0 or all equal to 1:
    $0<\sum_{i\in\mathcal{C}} w_{ij} < M,\ \forall j \in \mathcal{F}$.
    \item The number of ones in each column is at most $M/2$: $\sum_{i \in \mathcal{C}} w_{ij} \leq M / 2$.
    \item The code words of all categories are mutually distinct.
    \item All binary classifiers are distinct.
\end{itemize}

The Hamming distance between categories $c_i$ and $c_k$ is
\begin{equation}\label{eq:d_h_c}
    d_H(c_i,c_k) = \sum_{j = 1}^N w_{ij} \oplus w_{kj},
\end{equation}
where $\oplus$ denotes the exclusive OR; distinct code words give $d_H(c_i,c_k)\geq 1$ for $i\neq k$.
We write $d_{\min}\triangleq\min_{i\neq k}d_H(c_i,c_k)$ for the minimum distance of the table and, once the MAP category $c_1$ is fixed in Section~\ref{sec:PEP}, $d_{\min}(c_1)\triangleq\min_{k\neq1}d_H(c_1,c_k)\geq d_{\min}$ for the distance from its code word to the nearest other code word; the two coincide for equidistant codes.

\begin{definition}[$n$vR code]\label{def:nvR}
An $n$vR code is a code word table that contains all combinations partitioning the $M$ categories into $n$ versus $(M-n)$, $0<n\leq\lfloor M/2\rfloor$; for $n=M/2$ each partition is included once.
The Hamming distance of an $n$vR code is
\begin{align}\label{eq:d_nvR}
    d(n,M) = \begin{cases}
    2\binom{M - 2}{n - 1}, & n \neq  \frac{M}{2},\\
    \binom{M - 2}{n - 1}, & n =  \frac{M}{2}.
    \end{cases}
\end{align}
In particular, an $n$vR code is an equidistant code: $d_H(c_i,c_k)=d(n,M)$ for all $i\neq k$.
\end{definition}

\begin{definition}[Exhaustive Code]\label{def:exhaustive}
The Exhaustive Code is the code word table obtained by concatenating all $n$vR codes ($1\leq n\leq\lfloor M/2\rfloor$).
Its code length is $N_{\rm exh}=2^{M-1}-1$ and its minimum Hamming distance is $d_{\min}=2^{M-2}$.
\end{definition}

\begin{definition}[Hadamard code]\label{def:hadamard}
For $M\equiv0\pmod 4$ such that a Hadamard matrix of order $M$ exists (conjecturally every such $M$), the Hadamard code is obtained from a normalized Hadamard matrix of order $M$ by mapping $+1\mapsto0$ and $-1\mapsto1$; its $M$ code words of length $M$ are at mutual distance $M/2$ \citep{levenshtein1961hadamard,macwilliams1977theory}.
Its first column is constant, so as a code word table we delete it, which leaves an equidistant table with $N=M-1$ and $d=M/2$, attaining the Plotkin bound \citep{plotkin1960binary}; it is in general non-linear, and for $M=2^m$ with a Sylvester matrix it is the linear simplex code.
Equidistant codes of this kind have been constructed for ECOC from Reed--Muller codes and Hadamard matrices by \citet{hirasawa2022construction}; \citet{kumoi2022SMC} and \citet{hirasawa2022construction} call the code of length $M-1$ the simplex code. For $M=4$ it coincides with the 2vR code up to column complementation (Section~\ref{sec:specific}).
\end{definition}

The \emph{logM code}, whose code words are the binary representations of the category indices ($N=\lceil\log_2 M\rceil$, $d_{\min}=1$), is the shortest table and serves as the small-distance extreme in Section~\ref{sec:experiments}.

\subsection{Noise model and pairwise error probability}\label{sec:PEP}
We define the true posterior probability of each category for input data $\bm{x}$ by
$\theta_i = P(c_i|\bm{x})$
and write $\bm{\theta}=(\theta_1,\ldots,\theta_M)$.
Hereafter, $f_j(y|\bm{\theta})$ denotes the conditional probability of the binary category $y\in\{0,1\}$ given $\bm{x}$.
The output of a noiseless binary classifier $f_j^*$ is
\begin{equation}\label{eq:MAPecoc_nonnoise}
f^*_j(1|\bm{\theta}) =\sum_{i' \in \mathcal{C} : w_{i'j} = 1} \theta_{i'}.
\end{equation}
We use decoding based on the product of the posterior probabilities output by the binary classifiers,
\begin{equation}\label{eq:Passerini}
g(c_i|\bm{\theta}) = \prod_{j = 1}^N f_j(w_{ij}|\bm{\theta}),
\qquad \hat{i} = \argmax_{i \in \mathcal{C}} g(c_i|\bm{\theta}),
\end{equation}
and write $g^*(c_i|\bm{\theta}) = \prod_{j = 1}^N f_j^*(w_{ij}|\bm{\theta})$ in the noiseless case.

\begin{definition}\label{def:ECOC_noise}
The error term $\varepsilon_j$ of the log posterior probability ratio is
\begin{align}\label{eq:ecoc_noise}
\varepsilon_j = \ln \frac{f_j(1|\bm{\theta})}{1-f_j(1|\bm{\theta})} - \ln \frac{f^*_j(1|\bm{\theta})}{1-f^*_j(1|\bm{\theta})}.
\end{align}
If $f^*_j(1|\bm{\theta})\in\{0,1\}$, we define $f_j(1|\bm{\theta})=f^*_j(1|\bm{\theta})$ and $\varepsilon_j = 0$.
\end{definition}
Equivalently,
\begin{align}\label{eq:ecoc}
    f_j(1|\bm{\theta}) &= \frac{e^{\varepsilon_j} f^*_j(1|\bm{\theta})}{1 + (e^{\varepsilon_j} - 1) f^*_j(1|\bm{\theta})}.
\end{align}

\begin{assumption}\label{ass:eps}
$\varepsilon_j \sim \mathcal{N}(0,\sigma^2)$, i.i.d.
\end{assumption}

From \eqref{eq:Passerini} and \eqref{eq:ecoc}, the ratio of the decoding scores of $c_i$ and $c_k$ is the random variable
\begin{align}\label{eq:g_comp}
R_{ik} =\frac{g(c_i|\bm{\theta})}{g(c_k|\bm{\theta})} = e^{\sum_{j = 1}^N (w_{ij} - w_{kj})\varepsilon_j}\frac{g^*(c_i|\bm{\theta})}{g^*(c_k|\bm{\theta})}.
\end{align}

\begin{proposition}[Pairwise error probability \citep{kumoi2023IJNS}]\label{th:err_prob}
Under Assumption \ref{ass:eps},
\begin{align}\label{eq:err_prob}
    P(R_{ik} \leq 1|\bm{\theta})
    =\Phi\!\left(\frac{\ln\frac{g^*(c_k|\bm{\theta})}{g^*(c_i|\bm{\theta})}}{\sigma\sqrt{d_H(c_i,c_k)}}\right),
\end{align}
where $\Phi(\cdot)$ is the cumulative distribution function of the standard normal distribution.
\end{proposition}

We define
\begin{align}\label{eq:mu_rho}
\mu_{ik} &\triangleq \ln\frac{g^*(c_k|\bm{\theta})}{g^*(c_i|\bm{\theta})},\qquad
\rho_{ik} \triangleq \frac{-\mu_{ik}}{\sigma\sqrt{d_H(c_i,c_k)}}.
\end{align}
Hereafter the categories are ordered, without loss of generality, so that $\theta_1 > \theta_2 \geq \cdots \geq \theta_M$, and we assume $\theta_M > 0$; we work on the constraint set
\begin{align}\label{eq:Omega}
    \Omega(\lambda)=\{\bm{\theta}:\theta_1\geq\max_{k\neq 1} \theta_k + \lambda\}
\end{align}
with $\lambda>0$, on which the MAP category is $c_1$.
From Proposition \ref{th:err_prob} it follows that
\begin{align}\label{eq:PEP_Q}
P(R_{1k} \leq 1|\bm{\theta}) = \Phi(-\rho_{1k}) = Q(\rho_{1k}),
\end{align}
where $Q(x) = 1-\Phi(x)$; $\rho_{1k}\geq0$ when $\mu_{1k}\leq 0$, which holds for the $n$vR codes under $\bm{\theta}\in\Omega(\lambda)$ (Proposition~\ref{pro:mu_upper_nvR}).
Unless stated otherwise, Assumption~\ref{ass:eps} is in force in Sections~\ref{sec:channel}--\ref{sec:main_theorem}.

\section{Equivalent Channel and the Excess Error Rate}\label{sec:channel}

\subsection{Equivalent Gaussian channel}
From \eqref{eq:ecoc},
\begin{align}\label{eq:log_g}
    \ln g(c_i|\bm{\theta})
    &= \sum_{j=1}^N \Bigl[\ln f^*_j(w_{ij}|\bm{\theta})
      + w_{ij}\varepsilon_j - \ln\bigl(1 + (e^{\varepsilon_j}-1)f^*_j(1|\bm{\theta})\bigr)\Bigr].
\end{align}
The third term does not depend on $i$, so the decision of ECOC depends only on the sufficient statistic
\begin{equation}\label{eq:sufficient_stat}
    L_i \triangleq \sum_{j=1}^N w_{ij}\varepsilon_j + \ln g^*(c_i|\bm{\theta}),
    \qquad \hat{i} = \argmax_{i \in \mathcal{C}} L_i .
\end{equation}
Relative to the MAP category $c_1$,
\begin{equation}\label{eq:channel_model}
    L_k - L_1 = \mu_{1k} + Z_{1k},\qquad
    Z_{1k} \triangleq \sum_{j=1}^N (w_{kj} - w_{1j})\varepsilon_j \sim \mathcal{N}(0, d_H(c_1,c_k)\sigma^2),\quad k \neq 1 .
\end{equation}
This is an $M$-ary hypothesis test over a Gaussian channel: the transmitted word is the code word of $c_1$, the competitor $c_k$ is separated from it by the mean shift (signal) $\mu_{1k}$ and the distance $d_H(c_1,c_k)$, and $\rho_{1k}$ in \eqref{eq:mu_rho} is the corresponding signal-to-noise ratio.

\begin{remark}[Correspondence with channel coding]\label{rem:channel}
Unlike a standard AWGN channel, the signal $\mu_{1k}$ and the noise variance depend on the code-word pair through the Hamming distance.
For an equidistant code this dependence disappears, and each pairwise test $L_k-L_1$ is a standard binary AWGN decision.
The $\{Z_{1k}\}_{k\neq1}$ remain jointly dependent because they are built from the same $\{\varepsilon_j\}$, so the $M$-ary problem is not a set of independent parallel channels.
\end{remark}

\subsection{The excess error rate}\label{sec:EMDR}
We define the event that ECOC selects a category other than $c_1$ by
\begin{align}\label{eq:E}
    E = \bigcup_{k\neq1}\{R_{1k} \leq 1\},
\end{align}
so that $P(E|\bm{\theta})$ is the probability that ECOC fails to select the MAP category.
$E$ also contains the ties $R_{1k}=1$; under Assumption~\ref{ass:eps} (and under the correlated noise of Section~\ref{subsec:correlated}) every $\ln R_{1k}$ is a nondegenerate Gaussian, so ties have probability zero and $P(E|\bm{\theta})=P(\hat{i}_{\rm ECOC}\neq1|\bm{\theta})$.
We use this identity in \eqref{eq:P_ECOC_decomp} below and in the proof of Proposition~\ref{prop:Delta_basic}.
Let the true category be the random variable $C$ (with $P(C=c_i|\bm{\theta})=\theta_i$); given $\bm{\theta}$, the noise $\{\varepsilon_j\}$ is independent of $C$.
The MAP error rate is $P_{\rm MAP}(\bm{\theta}) = 1-\theta_1$, and the error rate of ECOC decomposes as
\begin{align}\label{eq:P_ECOC_decomp}
    P_{\rm ECOC}(\bm{\theta})
    &= \theta_1 P(E|\bm{\theta})
       + \sum_{k \neq 1} \theta_k P(\hat{i}_{\rm ECOC} \neq k \mid \bm{\theta}).
\end{align}

\begin{definition}[Excess error rate]
The (conditional) excess error rate of ECOC at $\bm{\theta}$ is
\begin{align}\label{eq:Delta_def}
    \Delta(\bm{\theta}) = P_{\rm ECOC}(\bm{\theta}) - P_{\rm MAP}(\bm{\theta}).
\end{align}
\end{definition}
Since $P_{\rm MAP}$ is the Bayes error rate at $\bm{x}$, $\Delta(\bm{\theta})$ is the conditional excess risk of ECOC under the 0--1 loss; averaging it over the input distribution gives the usual excess risk \citep{bartlett2006convexity,tewari2007consistency}.
Substituting $1-\theta_1=\sum_{k\neq 1}\theta_k$ into \eqref{eq:P_ECOC_decomp} gives the exact decomposition
\begin{align}\label{eq:Delta_exact}
    \Delta(\bm{\theta})
    &= \theta_1 P(E|\bm{\theta})
    - \sum_{k\neq 1}\theta_k P(\hat{i}_{\rm ECOC} = k|\bm{\theta}).
\end{align}

\begin{proposition}[Two-sided evaluation of the excess error rate]\label{prop:Delta_basic}
For any code word table and any $\bm{\theta}\in\Omega(\lambda)$, under any noise for which ties have probability zero (in particular under Assumption~\ref{ass:eps} and under the noise of Section~\ref{subsec:correlated}),
\begin{equation}\label{eq:Delta_basic_bounds}
    (\theta_1-\theta_2)\,P(E|\bm{\theta}) \leq \Delta(\bm{\theta}) \leq \theta_1 P(E|\bm{\theta}).
\end{equation}
In particular, since $\theta_1-\theta_2\geq\lambda$, we have $\Delta(\bm{\theta})\geq\lambda\,P(E|\bm{\theta})$.
\end{proposition}
\begin{proof}
\textit{Upper bound.} The second term of \eqref{eq:Delta_exact} is nonnegative, so $\Delta(\bm{\theta})\leq\theta_1 P(E|\bm{\theta})$.

\textit{Lower bound.} The events $\{\hat{i}_{\rm ECOC}=k\}$ ($k\neq 1$) are mutually disjoint, and $\{\hat{i}_{\rm ECOC}=k\}\subseteq\{R_{1k}\leq 1\}\subseteq E$, so
\begin{align}
    \sum_{k\neq 1}P(\hat{i}_{\rm ECOC}=k|\bm{\theta})
    = P(\hat{i}_{\rm ECOC}\neq 1|\bm{\theta})
    \leq P(E|\bm{\theta}).
\end{align}
Together with $\theta_k\leq\theta_2$ ($k\neq 1$),
\begin{align}
    \sum_{k\neq 1}\theta_k P(\hat{i}_{\rm ECOC}=k|\bm{\theta})
    \leq \theta_2\,P(E|\bm{\theta}).
\end{align}
Substituting this into \eqref{eq:Delta_exact} yields the lower bound.
\end{proof}

Thus $\Delta(\bm{\theta})$ and $P(E|\bm{\theta})$ differ by at most a factor in $[\lambda,\theta_1]$, and their normalized exponents differ by at most $(-\ln\lambda)/d_{\min}(c_1)$; Sections~\ref{sec:upper}--\ref{sec:tight} bound $P(E|\bm{\theta})$.

\subsection{The goal: a reliability function}

\begin{figure}[tb]
\centering
\includegraphics[width=\linewidth]{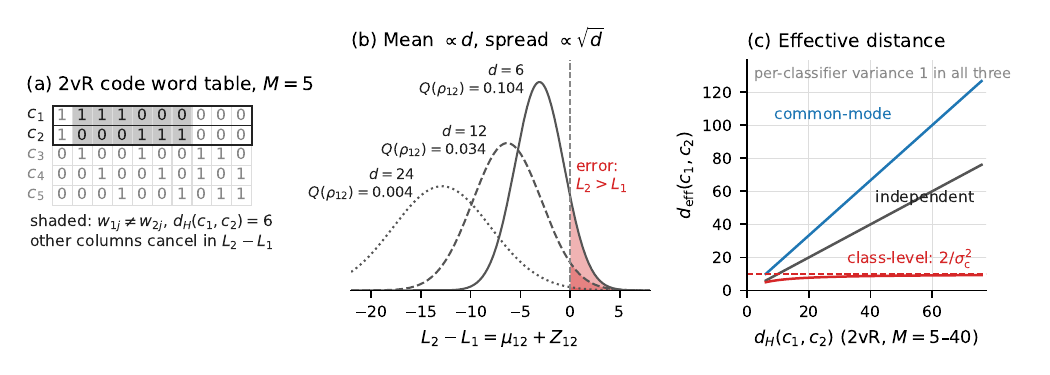}
\caption{Why the minimum distance governs the excess error rate.
(a)~In the 2vR code word table ($M=5$), only the $d_H(c_1,c_2)=6$ columns in which the code words of $c_1$ and $c_2$ differ enter the statistic $L_2-L_1$ of \eqref{eq:channel_model}; the other columns add the same amount to both scores.
(b)~The pairwise test $L_2-L_1=\mu_{12}+Z_{12}$ for 2vR codes of distance $d=6,12,24$ ($M=5,8,14$; $\theta_1=0.5$, $\theta_2=0.25$, the rest uniform, $\sigma=1$): the mean grows linearly in $d$ and the standard deviation as $\sqrt{d}$, so the error probability $Q(\rho_{12})$ (red) decays exponentially in $d$.
(c)~The effective distance of Section~\ref{subsec:correlated} along the 2vR family, for three noise models with the same variance $1$ of every classifier: independent ($\sigma^2=1$), common-mode ($\sigma_0^2=0.6$, $\sigma_{\rm z}^2=0.4$) and class-level ($\sigma_0^2=0.6$, $\sigma_{\rm c}^2=0.2$). Only class-level noise makes $d_{\rm eff}$ saturate, at $2/\sigma_{\rm c}^2$.}
\label{fig:concept}
\end{figure}

In channel coding, the reliability function is the exponent $E$ at which the error rate decays as $\exp(-N\cdot E)$ with the code length $N$.
Here the asymptotic variable is the minimum distance rather than $N$.
By \eqref{eq:channel_model}, the statistic $L_k-L_1$ depends only on the positions at which $c_1$ and $c_k$ differ.
Hence both the signal $\mu_{1k}$ and the noise variance $d_H(c_1,c_k)\sigma^2$ scale with $d_H$, and the effective code length for a pair is $d_H$ rather than $N$.
The error is then dominated by the nearest competitor, so the exponent of $P(E|\bm{\theta})$ is proportional to $d_{\min}(c_1)$, just as the minimum distance governs error correction in coding theory.
Since $N$ and $d_{\min}$ differ greatly across code families (for 2vR, $N=O(M^2)$ but $d_{\min}=O(M)$; for the logM code, $N=O(\log M)$ but $d_{\min}=1$; Table~\ref{tab:comparison}), the minimum distance is the natural variable.
Figure~\ref{fig:concept}(a,b) illustrates this mechanism.
Our goal is the exact exponent $E_{\rm ECOC}$ with $P(E|\bm{\theta})\asymp\exp(-d_{\min}(c_1)\cdot E_{\rm ECOC})$; for equidistant codes we write $d$ for the common distance.

\section{Upper Bound: Sufficiency of the Minimum Distance}\label{sec:upper}

\textbf{Strategy.} We bound $\mu_{1k}$ from above (Propositions~\ref{pro:mu_upper_general} and \ref{pro:mu_upper_nvR}), which bounds $\rho_{1k}$ from below (e.g.\ $\rho_{1k}\geq\rho^{\rm L}$, Eq.~\eqref{eq:rho_L}); since $Q$ is monotone decreasing, the union bound gives $P(E|\bm{\theta})\leq(M-1)Q(\rho^{\rm L})$, which Proposition~\ref{prop:Delta_basic} turns into an upper bound on the excess error rate (Proposition~\ref{prop:upper_general}, Corollary~\ref{cor:upper_nvR}).

\subsection{Upper bound on $\mu_{1k}$}

We introduce
\begin{align}\label{eq:omega_n}
    \omega(n) \triangleq \begin{cases}
        0, & n = 1,\\
        \sum_{k = M - (n - 2)}^M \theta_k, & n > 1,
    \end{cases}
    \qquad
    n_{\rm min} \triangleq \min_{j \in \mathcal{F} : w_{1j}=1} \sum_{i\in\mathcal{C}} w_{ij},
\end{align}
where $\omega(n)$ is the sum of the $n-1$ smallest posteriors and $n_{\rm min}$ is the minimum number of ones over the columns on whose positive side $c_1$ lies, and
\begin{align}\label{eq:alpha}
   \alpha_j \triangleq \sum_{i' \in \{\mathcal{C}\setminus\{1\} :  w_{1j} = w_{i'j}\}} \theta_{i'}.
\end{align}
The definition of $n_{\rm min}$ requires row $1$ of $W$ not to be identically $0$, which holds for every $n$vR code with $n<M/2$ (for $n$vR codes we use Corollary~\ref{cor:upper_nvR}, which does not involve $n_{\rm min}$) and is an additional assumption for a general code word table.

\begin{proposition}[Upper bound on $\mu_{1k}$: general code word table]\label{pro:mu_upper_general}
For $\bm{\theta}\in\Omega(\lambda)$,
\begin{align}\label{eq:mu_upper_general}
    \mu_{1k} \leq d_H(c_1,c_k)\ln\!\left(\frac{1-\theta_1-\omega(n_{\rm min})}{\theta_1+\omega(n_{\rm min})}\right).
\end{align}
\end{proposition}
\begin{proof}
From \eqref{eq:MAPecoc_nonnoise}, $f^*_j(w_{1j}|\bm{\theta})=\theta_1+\alpha_j$.
We show that $\alpha_j\geq\omega(n_{\rm min})$ holds for all $j\in\mathcal{F}$.
In a column with $w_{1j}=1$, $\alpha_j$ is the sum of the posterior probabilities of the $(n_j-1)$ categories other than $c_1$ on the positive side (with $n_j\geq n_{\rm min}$), and is at least the sum $\omega(n_{\rm min})$ of the $(n_{\rm min}-1)$ smallest posterior probabilities.
In a column with $w_{1j}=0$, $\alpha_j$ is the sum of the posterior probabilities of the $(M-n_j-1)$ categories other than $c_1$ on the negative side; by the properties of the code word table (the number of ones in each column is at most $M/2$), it contains $M-n_j-1\geq M/2-1\geq n_{\rm min}-1$ terms and is therefore again at least $\omega(n_{\rm min})$.
Since
\begin{align}\label{eq:ratio_bound}
    \frac{f^*_j(w_{kj}|\bm{\theta})}{f^*_j(w_{1j}|\bm{\theta})}
    =\frac{1-\theta_1-\alpha_j}{\theta_1+\alpha_j}
    \leq\frac{1-\theta_1-\omega(n_{\rm min})}{\theta_1+\omega(n_{\rm min})}
\end{align}
for $w_{1j}\neq w_{kj}$ and the left-hand side of \eqref{eq:ratio_bound} equals one for $w_{1j}=w_{kj}$, substituting these into
$\mu_{1k}=\sum_{j:w_{1j}\neq w_{kj}}\ln\frac{1-\theta_1-\alpha_j}{\theta_1+\alpha_j}$
yields \eqref{eq:mu_upper_general}.
\end{proof}

For $n$vR codes, a sharper bound follows from the pair structure.
For a pair of classifiers $f_j$ and $f_{j'}$ satisfying $w_{1j}=w_{kj'}\neq w_{kj}=w_{1j'}$ and $w_{lj}=w_{lj'}$ for all $l\neq 1,k$, define
$\eta_j(\alpha_j;\theta_1,\theta_k)\triangleq\frac{f^*_j(w_{kj}|\bm{\theta})}{f^*_j(w_{1j}|\bm{\theta})}\cdot\frac{f^*_{j'}(w_{kj'}|\bm{\theta})}{f^*_{j'}(w_{1j'}|\bm{\theta})}$.

\begin{lemma}[Upper bound on $\eta_j$]\label{lem:eta_upper}
If $\theta_1>\theta_k$, then on the domain $\alpha_j\in[0,1-\theta_1-\theta_k]$ (nonempty since $\theta_1+\theta_k\leq1$)
\begin{align}\label{eq:eta_upper}
   \eta_j(\alpha_j;\theta_1,\theta_k) \leq \left(\frac{1-\theta_1+\theta_k}{1+\theta_1-\theta_k}\right)^2,
\end{align}
with equality at $\alpha_j=(1-\theta_1-\theta_k)/2$.
\end{lemma}
As a function of $\alpha_j$ alone, $\eta_j$ is unimodal with its maximum at the interior point $\alpha_j=(1-\theta_1-\theta_k)/2$ (Appendix \ref{app:eta_upper}).

\begin{proposition}[Upper bound on $\mu_{1k}$: $n$vR code]\label{pro:mu_upper_nvR}
When the code word table is an $n$vR code,
\begin{align}\label{eq:mu_upper_nvR}
    \mu_{1k} \leq d(n,M)\ln\!\left(\frac{1-\theta_1+\theta_k}{1+\theta_1-\theta_k}\right).
\end{align}
\end{proposition}
\begin{proof}
Since an $n$vR code contains all partitions into $n$ versus $(M-n)$, for any $f_j$ with $w_{1j}\neq w_{kj}$ there is also a partition in the code word table that swaps only the positions of $c_1$ and $c_k$ while keeping the memberships of the other categories; this provides the counterpart $f_{j'}$ specified in Lemma \ref{lem:eta_upper}.
For $n=M/2$, a column separating $c_1$ and $c_k$ is determined by the set $\mathcal{S}$ of the other $n-1$ categories on the side of $c_1$, and the partition with $c_1$ and $c_k$ swapped corresponds to $\mathcal{R}\setminus\mathcal{S}$, where $\mathcal{R}=\mathcal{C}\setminus\{1,k\}$; this partition may be stored as its complement.
Complementing a column leaves the ratio $f^*_j(w_{kj}|\bm{\theta})/f^*_j(w_{1j}|\bm{\theta})$ unchanged, and the map $\mathcal{S}\mapsto\mathcal{R}\setminus\mathcal{S}$ has no fixed point for $n\geq2$, so the pairing argument is unaffected; the same applies to the proof of Proposition~\ref{pro:mu_lower_nvR}.
By equidistance, there are $d(n,M)$ columns with $w_{1j}\neq w_{kj}$, forming $d(n,M)/2$ such pairs.
By Lemma \ref{lem:eta_upper}, the contribution $\ln\eta_j$ of each pair is at most $2\ln\frac{1-\theta_1+\theta_k}{1+\theta_1-\theta_k}$, so
\begin{align}
    \mu_{1k}=\sum_{\text{pairs}}\ln\eta_j
    \leq \frac{d(n,M)}{2}\cdot 2\ln\frac{1-\theta_1+\theta_k}{1+\theta_1-\theta_k},
\end{align}
which gives \eqref{eq:mu_upper_nvR}.
\end{proof}

\subsection{Upper bound on the excess error rate}

\begin{proposition}[Upper bound: general code word table]\label{prop:upper_general}
For $\bm{\theta}\in\Omega(\lambda)$ with $\theta_1+\omega(n_{\rm min})>1/2$,
\begin{align}\label{eq:EMDR_upper_general}
\Delta(\bm{\theta})
\leq (M-1)\exp\!\left(-d_{\rm min}(c_1)\cdot \kappa_1(\theta_1, \omega(n_{\rm min}))\right),
\end{align}
where
$\kappa_1(\theta_1,\omega)\triangleq\frac{1}{2\sigma^2}\ln^2\!\left(\frac{\theta_1+\omega}{1-\theta_1-\omega}\right)$.
The bound also holds with $d_{\min}(c_1)$ replaced by $d_{\min}\leq d_{\min}(c_1)$.
\end{proposition}
\begin{proof}
Union bound: $P(E|\bm{\theta})\leq\sum_{k\neq 1}Q(\rho_{1k})$.
From Proposition \ref{pro:mu_upper_general} and the definition \eqref{eq:mu_rho} of $\rho_{1k}$,
\begin{align}
    \rho_{1k}\geq\frac{\sqrt{d_H(c_1,c_k)}}{\sigma}\ln\!\left(\frac{\theta_1+\omega(n_{\rm min})}{1-\theta_1-\omega(n_{\rm min})}\right).
\end{align}
When $\theta_1+\omega(n_{\rm min})>1/2$ we have $\rho_{1k}>0$, so applying the Chernoff bound $Q(x)\leq e^{-x^2/2}$ and using $d_H(c_1,c_k)\geq d_{\rm min}(c_1)\geq d_{\min}$ yields \eqref{eq:EMDR_upper_general}.
\end{proof}

\begin{corollary}[Upper bound: $n$vR code]\label{cor:upper_nvR}
For $\bm{\theta}\in\Omega(\lambda)$,
\begin{align}\label{eq:EMDR_upper_nvR}
\Delta(\bm{\theta})
\leq (M-1)\exp\!\left(-d(n,M)\cdot \kappa_2(\lambda,\sigma)\right),
\qquad
\kappa_2(\lambda,\sigma)\triangleq\frac{1}{2\sigma^2}\ln^2\!\left(\frac{1+\lambda}{1-\lambda}\right).
\end{align}
\end{corollary}
\begin{proof}
In Proposition \ref{pro:mu_upper_nvR}, using the monotone increase of $\ln\frac{1+u}{1-u}$ ($u=\theta_1-\theta_k$) and $\theta_1-\theta_k\geq\lambda$,
\begin{align}\label{eq:rho_L}
    \rho_{1k}\geq\frac{\sqrt{d(n,M)}}{\sigma}\ln\!\left(\frac{1+\lambda}{1-\lambda}\right)
    \triangleq\rho^{\rm L}.
\end{align}
Applying the union bound and the Chernoff bound yields \eqref{eq:EMDR_upper_nvR}.
\end{proof}

The union bound mirrors Gallager's method \citep{Gallager1968}: the excess error rate vanishes as soon as the distance term $\kappa\,d_{\min}(c_1)$ outgrows the logarithm $\ln(M-1)$ of the number of competitors, which plays the role of the rate.
Whether the exponent is exact requires the lower bound (Section~\ref{sec:lower}) and the tightness argument (Section~\ref{sec:tight}).

\section{Lower Bound: Necessity of the Hamming Distance}\label{sec:lower}

\textbf{Strategy.} In the equivalent channel, the error of the comparison with the hardest competitor lower-bounds $P(E|\bm{\theta})$ (Lemma~\ref{th:PE_lower}); this holds for any code word table and, with a column-wise bound on $\mu_{1k}$, yields necessity at once (Corollary~\ref{cor:necessity}). For $n$vR codes with $n\geq2$, a lower bound on $\mu_{1k}$ (Proposition~\ref{pro:mu_lower_nvR}) gives an explicit lower bound of the form $Q(\rho^{\rm U})$ (Proposition~\ref{th:EMDR_lower}).

\begin{lemma}[Two-sided evaluation of the Q function]\label{lem:Q_bounds}
For $x > 0$,
\begin{equation}\label{eq:Q_lower}
    \frac{1}{\sqrt{2\pi}}\cdot\frac{x}{x^2+1}\cdot e^{-x^2/2}
    \leq Q(x)
    \leq e^{-x^2/2},
\end{equation}
and for $x\geq 1$,
\begin{equation}\label{eq:Q_lower_simple}
    \frac{1}{2\sqrt{2\pi}\,x}\cdot e^{-x^2/2}
    \leq Q(x)
    \leq e^{-x^2/2}.
\end{equation}
\end{lemma}
\begin{proof}
The upper bound is the Chernoff bound.
The lower bound follows from the classical result on the Mills ratio \citep{Gordon1941,Feller1968}:
$Q(x)\geq\frac{1}{\sqrt{2\pi}}\cdot\frac{x}{x^2+1}\,e^{-x^2/2}$ ($x>0$), which gives \eqref{eq:Q_lower}.
For $x\geq 1$ we have $\frac{x}{x^2+1}\geq\frac{1}{2x}$ ($\Leftrightarrow 2x^2\geq x^2+1\Leftrightarrow x^2\geq 1$), so \eqref{eq:Q_lower} gives \eqref{eq:Q_lower_simple}.
\end{proof}

\subsection{Lower bound for any code word table and necessity}

Let
\begin{align}\label{eq:rho_min}
    \rho_{\min} \triangleq \min_{k\neq 1}\rho_{1k}.
\end{align}

\begin{lemma}[Lower bound on $P(E|\bm{\theta})$ (sphere-packing analogue)]\label{th:PE_lower}
Under Assumption \ref{ass:eps}, for any code word table,
\begin{equation}\label{eq:PE_lower}
    P(E|\bm{\theta}) \geq Q(\rho_{\min}),
\end{equation}
and, when $\rho_{\min}\geq 1$,
\begin{equation}\label{eq:PE_lower_mills}
    P(E|\bm{\theta})
    \geq \frac{1}{2\sqrt{2\pi}\,\rho_{\min}}\exp\!\left(-\frac{\rho_{\min}^2}{2}\right).
\end{equation}
\end{lemma}
\begin{proof}
From the definition \eqref{eq:E} of the event $E$, $\{R_{1k}\leq 1\}\subseteq E$ for any $k\neq 1$, so
\begin{align}
    P(E|\bm{\theta})
    \geq \max_{k\neq 1} P(R_{1k}\leq 1|\bm{\theta})
    = \max_{k\neq 1} Q(\rho_{1k})
    = Q(\rho_{\min}),
\end{align}
where the last equality follows from the monotone decrease of $Q$.
When $\rho_{\min}\geq 1$, applying the lower bound \eqref{eq:Q_lower_simple} of Lemma \ref{lem:Q_bounds} yields \eqref{eq:PE_lower_mills}.
\end{proof}
The bound plays the role of the sphere-packing bound in that the error against the nearest competitor cannot be avoided; unlike the sphere-packing bound of channel coding, it is a pairwise bound and involves no counting of decoding regions.

\begin{remark}[$\rho_{\min}$ for equidistant codes]\label{rem:kstar}
For an equidistant code, $d_H(c_1,c_k)$ does not depend on $k$, so $\rho_{1k}\propto|\mu_{1k}|$ and $\rho_{\min}$ is attained by the competitor with the smallest $|\mu_{1k}|$.
At finite $M$ this need not be $c_2$, but $\rho_{\min}\leq\rho_{12}$ always holds, so the lower bounds below are stated with $c_2$; for $n$vR codes, $\rho_{\min}=\rho_{12}$ for all sufficiently large $M$ in the asymptotic regime of Section~\ref{sec:tight} (Appendix~\ref{app:reliability}).
\end{remark}

\begin{corollary}[Necessity of the distance to the most probable competitor]\label{cor:necessity}
For $k\neq1$ let $B_k(\bm{\theta})\triangleq\ln\frac{1-\theta_k}{\theta_k}$, and let $B(\bm{\theta})\triangleq B_M(\bm{\theta})=\max_{k\neq1}B_k(\bm{\theta})$.
For any code word table and any $\bm{\theta}\in\Omega(\lambda)$,
\begin{equation}\label{eq:necessity}
    \Delta(\bm{\theta})
    \geq \lambda\,Q(\rho_{\min})
    \geq \lambda\,Q\!\left(\min_{k\neq1}\frac{\sqrt{d_H(c_1,c_k)}\,B_k(\bm{\theta})}{\sigma}\right)
    \geq \lambda\,Q\!\left(\frac{\sqrt{d_{\rm min}(c_1)}\,B(\bm{\theta})}{\sigma}\right) > 0.
\end{equation}
In particular $\Delta(\bm{\theta})\geq\lambda\,Q\bigl(\sqrt{d_H(c_1,c_2)}\,B_2(\bm{\theta})/\sigma\bigr)$, which depends on $\bm{\theta}$ only through $\lambda$ and $\theta_2$: as long as the distance between the code words of the two most probable categories is bounded, the excess error rate is bounded below by a positive constant that depends neither on the code length nor on $M$.
\end{corollary}
\begin{proof}
The first inequality follows from $\Delta(\bm{\theta})\geq\lambda P(E|\bm{\theta})$ (Proposition \ref{prop:Delta_basic}) and Lemma \ref{th:PE_lower}.
For the second, fix $k\neq1$ and a column $j$ with $w_{1j}\neq w_{kj}$.
The side containing $c_k$ has posterior mass $f^*_j(w_{kj}|\bm{\theta})\in[\theta_k,\,1-\theta_1]$ and the side containing $c_1$ has mass $f^*_j(w_{1j}|\bm{\theta})=1-f^*_j(w_{kj}|\bm{\theta})\in[\theta_1,\,1-\theta_k]$, so
\begin{align}
\frac{\theta_k}{1-\theta_k}\leq\frac{f^*_j(w_{kj}|\bm{\theta})}{f^*_j(w_{1j}|\bm{\theta})}\leq\frac{1-\theta_1}{\theta_1}.
\end{align}
Since $\theta_k<\theta_1$ and $\theta_1+\theta_k\leq1$, both $\bigl|\ln\frac{1-\theta_1}{\theta_1}\bigr|$ and $\ln\frac{1-\theta_k}{\theta_k}$ are at most $B_k(\bm{\theta})$, and the log ratio of every such column lies in $[-B_k(\bm{\theta}),B_k(\bm{\theta})]$.
Columns with $w_{1j}=w_{kj}$ do not contribute to $\mu_{1k}$, so
\begin{equation}\label{eq:mu_col_bound}
    |\mu_{1k}|\leq d_H(c_1,c_k)\,B_k(\bm{\theta})
\end{equation}
and
\begin{align}
    \rho_{\min}\leq\rho_{1k}\leq\frac{|\mu_{1k}|}{\sigma\sqrt{d_H(c_1,c_k)}}\leq\frac{\sqrt{d_H(c_1,c_k)}\,B_k(\bm{\theta})}{\sigma}
\end{align}
for every $k\neq1$; the second inequality of \eqref{eq:necessity} follows from the monotone decrease of $Q$.
The third follows by taking $k=\argmin_{k'\neq1}d_H(c_1,c_{k'})$ and using $B_k(\bm{\theta})\leq B_M(\bm{\theta})$, since $\theta_k\geq\theta_M$ and $\ln\frac{1-t}{t}$ is decreasing in $t$.
\end{proof}

\subsection{Explicit lower bound for $n$vR codes [$n\geq 2$]}\label{subsec:mu_lower_nvR}

For $n$vR codes with $n\geq 2$ we evaluate $\rho_{\min}$ as an explicit function of $\bm{\theta}$, which requires a lower bound on $\mu_{1k}$ (an upper bound on $|\mu_{1k}|$).

\begin{lemma}[Lower bound on $\eta$]\label{lem:eta_lower}
For the same pair $(f_j,f_{j'})$ as in Lemma \ref{lem:eta_upper}, if $\theta_1>\theta_k$ then
\begin{align}\label{eq:eta_lower}
   \eta_j(\alpha_j;\theta_1,\theta_k)
   \geq \frac{(1-\theta_1)\theta_k}{\theta_1(1-\theta_k)},
\end{align}
with equality at both endpoints $\alpha_j=0$ and $\alpha_j=1-\theta_1-\theta_k$.
\end{lemma}
\begin{proof}[Proof sketch]
When $\theta_1>\theta_k$, $\eta_j$ is unimodal in $\alpha_j$ on $[0,1-\theta_1-\theta_k]$ (increasing up to the interior maximum of Lemma~\ref{lem:eta_upper} and decreasing after it), so its minimum is attained at an endpoint, and both endpoint values equal $\frac{(1-\theta_1)\theta_k}{\theta_1(1-\theta_k)}$ (Appendix \ref{app:eta_lower}).
\end{proof}

\begin{proposition}[Lower bound on $\mu_{1k}$: $n$vR code, $n\geq 2$]\label{pro:mu_lower_nvR}
When the code word table is an $n$vR code,
\begin{align}\label{eq:mu_lower_nvR}
    \mu_{1k} \geq \frac{d(n,M)}{2}\ln\!\left(\frac{(1-\theta_1)\theta_k}{\theta_1(1-\theta_k)}\right).
\end{align}
\end{proposition}
\begin{proof}
From the equidistance and the pair structure of the $n$vR code,
\begin{align}
    \frac{g^*(c_k|\bm{\theta})}{g^*(c_1|\bm{\theta})}
    = \prod_{\text{pairs}} \eta_j(\alpha_j;\theta_1,\theta_k).
\end{align}
The number of pairs is $d(n,M)/2$, so applying Lemma \ref{lem:eta_lower},
\begin{align}
    \frac{g^*(c_k|\bm{\theta})}{g^*(c_1|\bm{\theta})}
    \geq \left(\frac{(1-\theta_1)\theta_k}{\theta_1(1-\theta_k)}\right)^{d(n,M)/2}.
\end{align}
Taking the logarithm yields \eqref{eq:mu_lower_nvR}.
\end{proof}

With
\begin{align}\label{eq:rho_U}
    \rho^{\rm U}_{1k} \triangleq \frac{\sqrt{d(n,M)}}{2\sigma}\left|\ln\frac{(1-\theta_1)\theta_k}{\theta_1(1-\theta_k)}\right|,
\end{align}
Proposition \ref{pro:mu_lower_nvR} gives $\rho_{1k}\leq\rho^{\rm U}_{1k}$.

\begin{proposition}[Lower bound on the excess error rate: $n$vR code, $n\geq 2$]\label{th:EMDR_lower}
When the code word table is an $n$vR code with $n\geq 2$, for $\bm{\theta}\in\Omega(\lambda)$,
\begin{equation}\label{eq:EMDR_lower_basic}
    \Delta(\bm{\theta})
    \geq \lambda\,Q(\rho^{\rm U}_{12}),
\end{equation}
and, when $\rho^{\rm U}_{12}\geq 1$,
\begin{align}\label{eq:EMDR_lower_refined}
    \Delta(\bm{\theta})
    \geq \frac{\lambda}{2\sqrt{2\pi}\,\rho^{\rm U}_{12}}
    \exp\!\left(-\frac{(\rho^{\rm U}_{12})^2}{2}\right),
    \qquad
    \frac{(\rho^{\rm U}_{12})^2}{2}
    = d(n,M)\cdot\frac{1}{8\sigma^2}\ln^2\!\left(\frac{\theta_1(1-\theta_2)}{(1-\theta_1)\theta_2}\right).
\end{align}
\end{proposition}
\begin{proof}
By Proposition \ref{prop:Delta_basic}, $\Delta(\bm{\theta})\geq\lambda P(E|\bm{\theta})$, and $\{R_{12}\leq1\}\subseteq E$ gives $P(E|\bm{\theta})\geq Q(\rho_{12})$.
From $\rho_{12}\leq\rho^{\rm U}_{12}$ and the monotone decrease of $Q$, we obtain \eqref{eq:EMDR_lower_basic}.
Equation \eqref{eq:EMDR_lower_refined} follows from the lower bound \eqref{eq:Q_lower_simple} of Lemma \ref{lem:Q_bounds}.
The expression for $(\rho^{\rm U}_{12})^2/2$ in \eqref{eq:EMDR_lower_refined} follows by substituting \eqref{eq:rho_U}.
\end{proof}

\section{Tightness: the Reliability Function}\label{sec:tight}

All results in this section are restricted to $n$vR codes with $n\geq 2$.

\textbf{Strategy.} For an equidistant code, Lemma~\ref{th:PE_lower} and the union bound give $Q(\rho_{12})\leq P(E|\bm{\theta})\leq(M-1)Q(\rho_{\min})$, and $\rho_{\min}=\rho_{12}$ for all sufficiently large $M$, so up to the prefactor the upper and lower exponents share the value $\rho_{12}^2/2$.
The tightness does not come from equating the bounds on $\mu_{1k}$ (their coefficients $d(n,M)$ and $d(n,M)/2$ differ); it follows from the sharp asymptotic value of $\mu_{12}$ (Lemma~\ref{lem:mu_asymptotic}), $\mu_{12}/(d(n,M)/2)\to\ln\frac{(1-\theta_1)\theta_2}{\theta_1(1-\theta_2)}$, so both exponents converge to
\begin{equation}\label{eq:elower_def}
e_{\rm lower}\triangleq\frac{1}{8\sigma^2}\ln^2\!\frac{\theta_1(1-\theta_2)}{(1-\theta_1)\theta_2}
\end{equation}
(Proposition~\ref{prop:reliability}); the subscript reflects that $e_{\rm lower}$ first appears as the exponent of the lower bound \eqref{eq:EMDR_lower_refined}.

\begin{lemma}[Sandwich on $P(E|\bm{\theta})$: $n$vR code, $n\geq 2$]\label{th:PE_nvR_sandwich}
When the code word table is an $n$vR code, for $\bm{\theta}\in\Omega(\lambda)$,
\begin{align}\label{eq:PE_nvR_lower}
    Q\!\left(\frac{\sqrt{d(n,M)}}{2\sigma}\left|\ln\frac{(1-\theta_1)\theta_2}{\theta_1(1-\theta_2)}\right|\right)
    \leq P(E|\bm{\theta})
    \leq (M-1)\exp\!\left(-\frac{d(n,M)}{2\sigma^2}\ln^2\!\left(\frac{1+\lambda}{1-\lambda}\right)\right).
\end{align}
\end{lemma}
\begin{proof}
Lower bound: by Proposition \ref{pro:mu_lower_nvR}, $\rho_{1k}\leq\rho^{\rm U}_{1k}$ (Eq.\ \eqref{eq:rho_U}).
From $P(E|\bm{\theta})\geq Q(\rho_{12})\geq Q(\rho^{\rm U}_{12})$ ($Q$ is monotone decreasing), we obtain the left inequality of \eqref{eq:PE_nvR_lower}.

Upper bound: the right inequality of \eqref{eq:PE_nvR_lower} follows from \eqref{eq:rho_L}, the union bound and the Chernoff bound, as in the proof of Corollary \ref{cor:upper_nvR}.
\end{proof}

\begin{remark}[Why the minimum distance suffices here]\label{rem:spectrum}
The factor $M-1$ in \eqref{eq:PE_nvR_lower} comes from a union bound over the $M-1$ competitors, each at distance $d$ --- exact for an equidistant code.
For equidistant signal sets on the additive white Gaussian noise channel, such as orthogonal and simplex signals, it is classical that this union bound and the pairwise lower bound share the exponent \citep{viterbi1979principles,Gallager1968}; here, unlike there, the signal $\mu_{1k}$ is set by the posterior rather than by a fixed signal energy and differs across competitors, and the exponent follows from the asymptotic value of $\mu_{12}$ (Lemma~\ref{lem:mu_asymptotic}).
For a general code the union bound is $\sum_{\delta} A_\delta\, Q(\rho(\delta))$, where $A_\delta$ is the number of competitors at distance $\delta$ from $c_1$, and $(M-1)Q(\rho_{\min})$ can be loose by up to a factor of $M$ when only one competitor is close.
The $n$vR restriction is thus the case in which the distance spectrum is degenerate and the minimum distance is a complete summary.
\end{remark}

\begin{lemma}[Asymptotically exact value of $\mu_{12}$: $n$vR code]\label{lem:mu_asymptotic}
For an $n$vR code with $n\geq 2$, fix $n$ and let $M\to\infty$.
Assume that $\theta_1,\theta_2$ are fixed independently of $M$ (with $1-\theta_1-\theta_2>0$) and that the posterior probabilities of the remaining categories satisfy $\theta_i=O(1/M)$ ($i\geq3$); the uniform allocation $\theta_i=(1-\theta_1-\theta_2)/(M-2)$ is an example. Then
\begin{equation}\label{eq:mu_asymptotic}
    \frac{\mu_{12}}{d(n,M)/2} \to \ln\frac{(1-\theta_1)\theta_2}{\theta_1(1-\theta_2)} \quad (M\to\infty),
\end{equation}
and, more precisely,
\begin{equation}\label{eq:mu_asymptotic_rate}
    \left|\frac{\mu_{12}}{d(n,M)/2} - \ln\frac{(1-\theta_1)\theta_2}{\theta_1(1-\theta_2)}\right|
    \leq \frac{\beta(\theta_1,\theta_2,n)}{M-2}
\end{equation}
for every $M\geq 2n$, where $\beta(\theta_1,\theta_2,n)\triangleq\bigl(\frac{1}{\theta_1}+\frac{1}{\theta_2}-\frac{1}{1-\theta_1}-\frac{1}{1-\theta_2}\bigr)(n-1)(1-\theta_1-\theta_2)$.
The bound \eqref{eq:mu_asymptotic_rate} uses only $\sum_{i\geq3}\theta_i=1-\theta_1-\theta_2$, not $\theta_i=O(1/M)$; if moreover $\theta_i>0$ for all $i\geq3$, then $\mu_{12}/(d(n,M)/2)$ is strictly larger than its limit for every $M\geq2n$.
\end{lemma}
\begin{proof}[Proof sketch]
The log ratio $\gamma(\alpha_{\mathcal{S}})=\ln\frac{(1-\theta_1-\alpha_{\mathcal{S}})(\theta_2+\alpha_{\mathcal{S}})}{(\theta_1+\alpha_{\mathcal{S}})(1-\theta_2-\alpha_{\mathcal{S}})}$ of each classifier pair satisfies $\gamma(0)\leq\gamma(\alpha_{\mathcal{S}})\leq\gamma(0)+c'\alpha_{\mathcal{S}}$ with $c'=1/\theta_1+1/\theta_2-1/(1-\theta_1)-1/(1-\theta_2)$; averaging over the subsets $\mathcal{S}$ spanning the pairs, whose mean $\bar{\alpha}_{\mathcal{S}}$ is $(n-1)(1-\theta_1-\theta_2)/(M-2)$, gives the claim (Appendix \ref{app:mu_asymptotic}).
\end{proof}

\begin{definition}[ECOC reliability function]\label{def:reliability}
For an $n$vR code with $n$ fixed and $M\to\infty$ (hence $d(n,M)\to\infty$),
\begin{equation}\label{eq:reliability_def}
    E_{\rm ECOC}(\bm{\theta},\sigma)
    \triangleq -\lim_{d(n,M)\to\infty}
    \frac{\ln P(E|\bm{\theta})}{d(n,M)}.
\end{equation}
Since $\bm{\theta}$ has dimension $M$, it is understood as a sequence $\bm{\theta}^{(M)}\in\Omega(\lambda)$ with $\theta_1^{(M)}=\theta_1$ and $\theta_2^{(M)}=\theta_2$ fixed and $\theta_i^{(M)}=O(1/M)$ ($i\geq3$), the family of Lemma \ref{lem:mu_asymptotic}; Proposition \ref{prop:reliability} shows that the limit does not depend on the particular sequence.
\end{definition}

\begin{proposition}[Existence and value of the reliability function]\label{prop:reliability}
For an $n$vR code with $n\geq 2$ and any sequence $\bm{\theta}^{(M)}$ of the family above, the ECOC reliability function exists and
\begin{equation}\label{eq:reliability_value}
    E_{\rm ECOC}(\bm{\theta},\sigma) = e_{\rm lower}=\frac{1}{8\sigma^2}
    \ln^2\!\left(\frac{\theta_1(1-\theta_2)}{(1-\theta_1)\theta_2}\right).
\end{equation}
\end{proposition}
\begin{proof}[Proof sketch]
By Lemma \ref{lem:mu_asymptotic}, $\rho_{12}^2/(2d(n,M))\to e_{\rm lower}$, and $\rho_{\min}=\rho_{12}$ for all sufficiently large $M$.
The union and Chernoff bounds give $\liminf\frac{-\ln P(E|\bm{\theta})}{d}\geq e_{\rm lower}$, and Lemmas \ref{th:PE_lower} and \ref{lem:Q_bounds} give $\limsup\frac{-\ln P(E|\bm{\theta})}{d}\leq e_{\rm lower}$, so the limit exists and equals $e_{\rm lower}$ (Appendix \ref{app:reliability}).
\end{proof}

\begin{corollary}[Reliability function of the excess error rate]\label{cor:EMDR_reliability}
For an $n$vR code with $n\geq 2$,
\begin{equation}\label{eq:EMDR_reliability}
    -\lim_{d(n,M)\to\infty}\frac{\ln\Delta(\bm{\theta})}{d(n,M)}
    = E_{\rm ECOC}(\bm{\theta},\sigma).
\end{equation}
\end{corollary}
\begin{proof}
Taking the logarithm of the two-sided evaluation
$\lambda P(E|\bm{\theta})\leq\Delta(\bm{\theta})\leq\theta_1 P(E|\bm{\theta})$
of Proposition \ref{prop:Delta_basic} and dividing by $d(n,M)$,
\begin{align}
    \frac{-\ln(\theta_1 P(E|\bm{\theta}))}{d(n,M)}
    \leq \frac{-\ln\Delta(\bm{\theta})}{d(n,M)}
    \leq \frac{-\ln(\lambda P(E|\bm{\theta}))}{d(n,M)}.
\end{align}
Since $\lambda,\theta_1$ are positive constants independent of $d(n,M)$, we have $-\ln\lambda/d(n,M)\to 0$ and $-\ln\theta_1/d(n,M)\to 0$ as $d(n,M)\to\infty$.
Hence both sides have the same limit as $-\ln P(E|\bm{\theta})/d(n,M)$, and \eqref{eq:EMDR_reliability} follows from Proposition \ref{prop:reliability}.
\end{proof}

The coefficient $\kappa_2(\lambda,\sigma)$ of Corollary \ref{cor:upper_nvR} holds uniformly over $\Omega(\lambda)$, whereas $E_{\rm ECOC}(\bm{\theta},\sigma)$ is the exact exponent at each $\bm{\theta}$; they are related as follows.

\begin{lemma}[Uniform and pointwise exponents]\label{lem:c2_elower}
For any $\bm{\theta}\in\Omega(\lambda)$, $E_{\rm ECOC}(\bm{\theta},\sigma) \geq \kappa_2(\lambda,\sigma)$,
with equality in the limit $\theta_1+\theta_2\to1$, $\theta_1-\theta_2=\lambda$; that is, $\kappa_2(\lambda,\sigma)=\inf_{\bm{\theta}\in\Omega(\lambda)}E_{\rm ECOC}(\bm{\theta},\sigma)$ (Appendix~\ref{app:c2_elower}).
\end{lemma}

The convergence of Lemma~\ref{lem:mu_asymptotic} (the $O(1/M)$ rate, $M=4$--$500$), that of the Monte Carlo exponent $-\ln P(E|\bm{\theta})/d(n,M)$ to $E_{\rm ECOC}$, and the consistency of Lemma~\ref{th:PE_nvR_sandwich} with Monte Carlo estimates ($4\leq M\leq20$) are verified numerically in Appendix~\ref{app:numerical}.
In all cases examined there ($n=2$), the finite-length exponent exceeds $e_{\rm lower}$.
This excess comes from the subexponential prefactor of $P(E|\bm{\theta})$; the correction term of Lemma~\ref{lem:mu_asymptotic} acts in the opposite direction, since $r(M)>0$ implies $\rho_{12}^2/(2d(n,M))<e_{\rm lower}$.
The excess is therefore not guaranteed in general: for $n\geq3$ and the family of Definition~\ref{def:reliability}, $d(n,M)=\Theta(M^{n-1})$ grows fast enough that the $O(1/M)$ correction dominates the prefactor, and Lemma~\ref{th:PE_lower} with \eqref{eq:Q_lower} gives $-\ln P(E|\bm{\theta})/d(n,M)<e_{\rm lower}$ for sufficiently large $M$ (in the setting of Appendix~\ref{app:numerical}, for $n=3$ and $M\geq25$).

\section{Main Theorem and Specific Codes}\label{sec:main_theorem}

\begin{theorem}[Characterization of the excess error rate by the minimum distance]\label{th:main}
Under Assumption \ref{ass:eps}, for $\bm{\theta}\in\Omega(\lambda)$ ($\lambda>0$):
\begin{enumerate}
    \item \textbf{Sufficiency} (Proposition \ref{prop:upper_general}, Corollary \ref{cor:upper_nvR}): $\Delta(\bm{\theta})\leq(M-1)\exp(-\kappa\,d_{\rm min}(c_1))$, with $\kappa=\kappa_1(\theta_1,\omega(n_{\rm min}))$ for any code word table satisfying the conditions of Proposition~\ref{prop:upper_general} and $\kappa=\kappa_2(\lambda,\sigma)$ for $n$vR codes. Since the binary classifiers are distinct, $N<2^M$, so $d_{\rm min}(c_1)\leq N$ is bounded at fixed $M$ and can grow only together with $M$; along any sequence of code word tables, $\Delta(\bm{\theta})\to 0$ whenever $\kappa\,d_{\min}(c_1)-\ln(M-1)\to\infty$. For $n$vR codes with $n\geq2$ this holds as $M\to\infty$, uniformly over $\Omega(\lambda)$, since $\ln(M-1)=o(d(n,M))$.
    \item \textbf{Necessity} (Corollary \ref{cor:necessity}): for any code word table, $\Delta(\bm{\theta})\geq\lambda\,Q\bigl(\sqrt{d_H(c_1,c_2)}\,B_2(\bm{\theta})/\sigma\bigr)>0$. Hence, if $d_H(c_1,c_2)$ (in particular, if $d_{\min}(c_1)$ for an equidistant code) stays bounded, the excess error rate is bounded below by a positive constant determined by $\lambda$, $\theta_2$, $\sigma$ and that distance, uniformly in $N$ and $M$.
    \item \textbf{Tightness} (Proposition \ref{prop:reliability}): for an $n$vR code with $n\geq 2$, the upper and lower exponents coincide, and the reliability function is $E_{\rm ECOC}(\bm{\theta},\sigma) = \frac{1}{8\sigma^2}\ln^2\!\left(\frac{\theta_1(1-\theta_2)}{(1-\theta_1)\theta_2}\right)$.
\end{enumerate}
\end{theorem}

Theorem~\ref{th:main} gives ECOC counterparts of the parts of the channel coding theorem for a fixed code: sufficiency to achievability, necessity to the converse, and tightness to the reliability function.
The corresponding statement for codes drawn at random is the average bound of \citet{kobayashi2024average}, in which the code length enters through the rate $\ln M/N$; in Theorem~\ref{th:main} the role of $N$ is taken by $d_{\min}(c_1)$.
This is the regime ECOC codes operate in: for $n$vR codes with $n\geq2$ and for Hadamard codes the rate $\ln M/N$ tends to zero, and at low rates the error exponent of channel codes is already governed by the probability of confusing two code words at the minimum distance rather than by the code length \citep[Section~V]{gallager1965simple}.
Necessity holds for every code word table, sufficiency for every code word table satisfying the conditions of Proposition~\ref{prop:upper_general}, and the value of the reliability function is proved for $n$vR codes with $n\geq2$ ($n$ fixed, $M\to\infty$).
Assumption~\ref{ass:eps} enters Theorem~\ref{th:main} only through the marginal law of each $Z_{1k}$ in \eqref{eq:channel_model}; Theorem~\ref{th:correlated} in Section~\ref{subsec:correlated} removes the independence.

\subsection{Specific codes}\label{sec:specific}

Table~\ref{tab:comparison} summarizes the orders of the exponent $-\ln\Delta(\bm{\theta})$ for fixed $\theta_1$ and $\theta_2$.
For every row, an upper order follows from Corollary~\ref{cor:necessity} with $k=2$.
For 2vR, Hadamard and Exhaustive, whose sufficiency bounds vanish, the matching lower order follows from the upper bounds of Section~\ref{sec:upper}. For 1vR and logM the minimum distance is bounded; since $\Delta(\bm{\theta})\leq\theta_1$, the order is $\Theta(1)$ whenever Corollary~\ref{cor:necessity} gives $O(1)$, and for logM with growing $d_H(c_1,c_2)$ only the upper order $O(\log M)$ is available.

\textbf{1vR and logM.} For 1vR, $d(1,M)=2$ and $\mu_{1k}=\ln\frac{(1-\theta_1)\theta_k}{\theta_1(1-\theta_k)}$ exactly ($\alpha_j=0$, the equality case of Lemma~\ref{lem:eta_lower}), so Corollary~\ref{cor:necessity} bounds the excess error rate away from zero uniformly in $M$, and the reliability function is not defined. For logM, $d_H(c_1,c_2)\leq N=\lceil\log_2M\rceil$, so Corollary~\ref{cor:necessity} gives $-\ln\Delta(\bm{\theta})=O(\log M)$, and the excess error rate stays bounded away from zero whenever the code words of $c_1$ and $c_2$ are at a bounded distance (for example, adjacent indices in the binary representation).

\textbf{2vR.} Since $d(2,M)=2(M-2)\to\infty$ (for $M\geq5$; $d(2,4)=2$ by \eqref{eq:d_nvR}), the excess error rate decays exponentially in $d(2,M)$, uniformly over $\Omega(\lambda)$ at rate $\kappa_2(\lambda,\sigma)$ (Corollary~\ref{cor:upper_nvR}) and, at each $\bm{\theta}$, at the exact rate $E_{\rm ECOC}(\bm{\theta},\sigma)\geq\kappa_2$ (Proposition~\ref{prop:reliability}).

\textbf{Hadamard.} The Hadamard code (Definition~\ref{def:hadamard}) is equidistant with $d=M/2$ at code length $N=M-1$, but for $M\geq8$ it has no pair structure in the sense of Lemma~\ref{lem:eta_upper}, so Propositions~\ref{pro:mu_upper_nvR} and \ref{pro:mu_lower_nvR} do not apply.
Indeed, in $\pm1$ form two paired columns differ exactly in rows $1$ and $k$, so their entrywise product has exactly $2$ entries equal to $-1$ ($M-2$ if one of them is complemented, which leaves every ratio $f^*_j(w_{kj}|\bm{\theta})/f^*_j(w_{1j}|\bm{\theta})$ unchanged), whereas distinct columns of a Hadamard matrix are orthogonal, so the product has exactly $M/2$ such entries.
The two counts agree only for $M=4$, where the Hadamard code is 2vR up to column complementation; for $M\geq8$ no category pair $(c_1,c_k)$ has even one paired column. Bounds of the form \eqref{eq:mu_upper_nvR} and \eqref{eq:mu_lower_nvR} may still hold for the Hadamard code by another argument; this is open.
Every column of the table has $M/2$ ones, and after complementing columns if necessary (which does not change the decoding problem), row $1$ is not identically $0$, so $n_{\min}=M/2$.
Proposition~\ref{prop:upper_general} then gives $\Delta(\bm{\theta})\leq(M-1)\exp(-\tfrac{M}{2}\kappa_1(\theta_1,\omega(M/2)))$ whenever $\theta_1+\omega(M/2)>1/2$.
Combined with Corollary~\ref{cor:necessity} for $k=2$, the exponent is of order $M$ at code length $O(M)$ when $\theta_1+\omega(M/2)\geq1/2+c$ for a constant $c>0$. For example, if the posterior mass outside $c_1$ and $c_2$ is spread uniformly, $\omega(M/2)=(1-\theta_1-\theta_2)/2$ and the condition holds with $c=\lambda/2$ for every $\bm{\theta}\in\Omega(\lambda)$; it can fail when that mass is concentrated on a few categories, where it reduces to $\theta_1\geq1/2+c$.
Whether it attains a reliability function, and its coefficient, are open.

\textbf{Exhaustive.} As the concatenation of all $n$vR codes it is equidistant with $d_{\rm exh}=\sum_{n=1}^{\lfloor M/2\rfloor}d(n,M)=2^{M-2}$.
Applying the $\mu_{1k}$ bounds block-wise, the excess error rate vanishes doubly exponentially in $M$, with exponent of order $N_{\rm exh}=2^{M-1}-1$. Only the order is tight (coefficients $\kappa_2$ versus the pointwise $e_{\rm exh}(\bm{\theta})\geq\kappa_2$, since $\bar\alpha_{\mathcal{S}}\to0$ in Lemma~\ref{lem:mu_asymptotic} fails for blocks with $n\sim M/2$); see Corollary~\ref{th:exhaustive_tight} (Appendix~\ref{app:exhaustive}).

\begin{table}[htbp]
\centering
\caption{Code length, minimum distance, and order of the exponent $-\ln\Delta(\bm{\theta})$ ($\theta_1,\theta_2$ fixed).}
\label{tab:comparison}
\begin{tabular}{l|c|c|c}
\hline
Code & $N$ & $d_{\rm min}$ & Order of $-\ln\Delta(\bm{\theta})$ \\\hline
logM & $\lceil\log_2 M\rceil$ & $1$ & $O(\log M)$; $\Theta(1)$ if $d_H(c_1,c_2)$ is bounded \\
1vR & $M$ & 2 & $\Theta(1)$ (no decay) \\
2vR & $M(M-1)/2$ & $2(M-2)$ & $\Theta(M)$; $-\ln\Delta(\bm{\theta})\sim 2M\,E_{\rm ECOC}$ \\
Hadamard & $M-1$ & $M/2$ & $\Theta(M)$ if $\theta_1+\omega(M/2)\geq1/2+c$; coefficient open \\
Exhaustive & $2^{M-1}-1$ & $2^{M-2}$ & $\Theta(2^M)$; coefficient open \\\hline
\end{tabular}

\vspace{2pt}
\begin{minipage}{0.9\linewidth}
\footnotesize $O(\log M)$ for logM is an upper order only; $c>0$ is a constant independent of $M$; the 2vR row assumes $M\geq5$ and its coefficient the posterior family of Definition~\ref{def:reliability}.
\end{minipage}
\end{table}

Among these, the Hadamard code is the only one reaching an exponent of order $M$ at code length $O(M)$ (ratio $d_{\min}/N\approx1/2$).

\section{Correlated Noise and the Effective Distance}\label{subsec:correlated}

Assumption~\ref{ass:eps} requires independent noise, but Sections~\ref{sec:channel}--\ref{sec:tight} use it only through the marginal law of each $Z_{1k}$ in \eqref{eq:channel_model}, a centered Gaussian with variance $d_H(c_1,c_k)\sigma^2$; the joint law of $\{Z_{1k}\}_{k\neq1}$ is never used (cf.\ Remark~\ref{rem:channel}).
Let $\bm{\varepsilon}=(\varepsilon_1,\ldots,\varepsilon_N)^\top\sim\mathcal{N}(\mathbf{0},\Sigma)$ with an arbitrary covariance $\Sigma\succ0$, and let $\bm{s}_k\triangleq(w_{kj}-w_{1j})_{j=1}^N$, so that $Z_{1k}=\bm{s}_k^\top\bm{\varepsilon}\sim\mathcal{N}(0,\bm{s}_k^\top\Sigma\bm{s}_k)$.
We define the effective noise level and the \emph{effective distance} of the pair $(c_1,c_k)$ by
\begin{equation}\label{eq:d_eff}
    \sigma^2_{\rm eff}(1,k)\triangleq\frac{\bm{s}_k^\top\Sigma\bm{s}_k}{d_H(c_1,c_k)},\qquad
    d_{\rm eff}(c_1,c_k)\triangleq\frac{d_H(c_1,c_k)}{\sigma^2_{\rm eff}(1,k)}=\frac{d_H(c_1,c_k)^2}{\bm{s}_k^\top\Sigma\bm{s}_k}.
\end{equation}
Under Assumption~\ref{ass:eps}, $\sigma_{\rm eff}(1,k)=\sigma$ and $d_{\rm eff}(c_1,c_k)=d_H(c_1,c_k)/\sigma^2$.

\begin{theorem}[Correlated Gaussian noise]\label{th:correlated}
Let $\bm{\varepsilon}\sim\mathcal{N}(\mathbf{0},\Sigma)$ with $\Sigma\succ0$ in place of Assumption~\ref{ass:eps}, and let $\bm{\theta}\in\Omega(\lambda)$ ($\lambda>0$).
\begin{enumerate}
    \item \textbf{Theorem~\ref{th:main} with the effective distance.} Parts~1 and~2 of Theorem~\ref{th:main} hold with $d_H(c_1,c_k)/\sigma^2$ replaced by $d_{\rm eff}(c_1,c_k)$:
    \begin{equation}\label{eq:correlated_bounds}
        \Delta(\bm{\theta})\leq(M-1)\exp\Bigl(-\bar\kappa\,\min_{k\neq1}d_{\rm eff}(c_1,c_k)\Bigr),\qquad
        \Delta(\bm{\theta})\geq\lambda\,Q\Bigl(\sqrt{d_{\rm eff}(c_1,c_2)}\,B_2(\bm{\theta})\Bigr)>0,
    \end{equation}
    where $\bar\kappa$ is the constant $\kappa$ of Theorem~\ref{th:main} evaluated at $\sigma=1$, under the same conditions.
    For an $n$vR code with $n\geq2$ and a sequence $\bm{\theta}^{(M)}$ of the family of Definition~\ref{def:reliability}, part~3 holds with $\sigma^2$ replaced by $v$, that is, $-\ln\Delta(\bm{\theta})/d(n,M)\to E_{\rm ECOC}(\bm{\theta},\sqrt{v})$, whenever $\sigma^2_{\rm eff}(1,2)\to v\in(0,\infty)$ and $\max_{k\neq1}\sigma^2_{\rm eff}(1,k)$ stays bounded as $M\to\infty$.
    \item \textbf{Common-mode versus class-level noise.} Let $\varepsilon_j=\xi_j+\sum_{i:\,w_{ij}=1}\chi_i+z$ with independent $\xi_j\sim\mathcal{N}(0,\sigma_0^2)$, $\chi_i\sim\mathcal{N}(0,\sigma_{\rm c}^2)$ and $z\sim\mathcal{N}(0,\sigma_{\rm z}^2)$, $\sigma_0>0$.
    If $\sigma_{\rm c}>0$, then for every code word table $d_{\rm eff}(c_1,c_k)\leq2/\sigma_{\rm c}^2$ for all $k\neq1$, and hence $\Delta(\bm{\theta})\geq\lambda\,Q\bigl(\sqrt{2}\,B_2(\bm{\theta})/\sigma_{\rm c}\bigr)>0$ whatever $N$ and $d_{\min}$.
    If $W$ is equidistant with distance $d$ and has equal row weights, $\hat{i}$ has the same distribution as under Assumption~\ref{ass:eps} with $\sigma^2=\sigma_0^2+\sigma_{\rm c}^2d/2$; if $\sigma_{\rm c}=0$, equal row weights alone suffice.
\end{enumerate}
\end{theorem}
\begin{proof}[Proof sketch]
(1)~With $\rho_{1k}\triangleq-\mu_{1k}/(\bm{s}_k^\top\Sigma\bm{s}_k)^{1/2}$, \eqref{eq:PEP_Q} holds as before; Proposition~\ref{prop:Delta_basic} and the bounds on $\mu_{1k}$ do not involve the noise, and the union bound and Lemma~\ref{th:PE_lower} use only the law of each $Z_{1k}$.
(2)~Here $\Sigma=\sigma_0^2I+\sigma_{\rm c}^2W^\top W+\sigma_{\rm z}^2\mathbf{1}\mathbf{1}^\top\succeq\sigma_{\rm c}^2W^\top W$, and $d_H(c_1,c_k)=(\bm{e}_k-\bm{e}_1)^\top W\bm{s}_k\leq\sqrt{2}\,\lVert W\bm{s}_k\rVert$, so $\bm{s}_k^\top\Sigma\bm{s}_k\geq\sigma_{\rm c}^2d_H(c_1,c_k)^2/2$.
The details are in Appendix~\ref{app:correlated}.
\end{proof}

The case $\sigma_{\rm c}=0$ is equicorrelated noise, $\mathrm{Cov}(\bm{\varepsilon})=\sigma^2[(1-\tau)I+\tau\mathbf{1}\mathbf{1}^\top]$ with $\sigma_0^2=(1-\tau)\sigma^2$ and $\sigma_{\rm z}^2=\tau\sigma^2$: correlation common to all classifiers cancels in every score difference and only lowers the effective noise level.
The $n$vR codes with $n<M/2$ have equal row weights $\binom{M-1}{n-1}$, so for them Theorem~\ref{th:main} applies with $(1-\tau)\sigma^2$.
Class-level noise behaves differently.
For $n$vR codes with $n<M/2$, every classifier has the same marginal variance $\sigma_0^2+n\sigma_{\rm c}^2+\sigma_{\rm z}^2$, independent of $M$, yet $d_{\rm eff}=d/(\sigma_0^2+\sigma_{\rm c}^2d/2)$ saturates at $2/\sigma_{\rm c}^2$ as $d(n,M)$ grows, and the excess error rate no longer vanishes (Figure~\ref{fig:concept}(c)).
The quantity that must outgrow $\ln(M-1)$ is therefore the effective distance, which combines the separation of the code words with the correlation of the classifier errors; the Hamming distance is its value under independent noise.
Theorem~\ref{th:correlated} concerns the decoder \eqref{eq:Passerini}. In part~2 the class components enter through the positive side of each column, but the floor does not depend on this orientation convention: if each class component enters with the sign of its side, $\varepsilon_j=\xi_j+\sum_i(2w_{ij}-1)\chi_i+z$, the same argument with $(\bm{e}_k-\bm{e}_1)^\top(2W-\mathbf{1}\mathbf{1}^\top)\bm{s}_k=2d_H(c_1,c_k)$ gives $d_{\rm eff}(c_1,c_k)\leq1/(2\sigma_{\rm c}^2)$.

\section{Experiments}\label{sec:experiments}

\subsection{Setup}\label{subsec:exp_setup}

We check two predictions: (i)~in a setting that satisfies Assumption~\ref{ass:eps}, the excess error rate decays exponentially with $d$ at the rate $e_{\rm lower}$; (ii)~on trained classifiers, the excess error rate decreases with the minimum distance and, beyond a moderate distance, no longer improves appreciably.
The real-data check is qualitative, because trained classifiers do not satisfy Assumption~\ref{ass:eps} exactly and no procedure for estimating their noise level stably without the true posterior has been established (Section~\ref{sec:discussion}).

\textbf{L0 (semi-synthetic) setting.}
Log-odds noise satisfying Assumption~\ref{ass:eps} with known $\sigma$ is added to the true $f^*_j$ of the 2vR code, with $\theta_1\in\{0.4,0.5,0.6,0.7\}$, $\theta_2=0.25$, the remaining mass spread uniformly, and $6\leq M\leq24$, so that $d=2(M-2)$ ranges from $8$ to $44$; $P(E|\bm{\theta})$ and $\Delta(\bm{\theta})$ are estimated by Monte Carlo.

\textbf{Real data.}
We use ResNet-50 features (pretrained on ImageNet-1k) of CIFAR-10, CIFAR-100 and ImageNet-1k, reduced by PCA to $100$ dimensions.
In each of $S=40$ seeds we draw $M$ classes ($M=8$, $12$ and $16$, respectively) and train on $400$ (CIFAR) or $150$ (ImageNet-1k) images per class; we test on the CIFAR test sets and the ImageNet-1k validation set.
Each binary classifier is a logistic regression (inverse regularization strength $1.0$) whose output $p_j$ estimates $f_j(1|\bm{x})$.
To vary $d_{\min}$ at fixed $M$, we draw subsets of $K$ columns from a fixed pool of 2vR and 3vR columns (all $84$ and $286$ columns for $M=8$ and $12$; the $120$ 2vR columns and $200$ randomly chosen 3vR columns for $M=16$), and record the minimum distance of each subset; the points with $d_{\min}=1$ are such subsets, not the logM code.
The ECOC prediction uses sum decoding $\hat{\imath}_{\rm sum}=\argmax_{i\in\mathcal{C}}\sum_j[w_{ij}p_j+(1-w_{ij})(1-p_j)]$.

\textbf{Reference error rate.}
On real data the Bayes error rate is unknown.
As a reference we use the error rate $P_{\rm floor}$ of log-likelihood (LL) decoding $\hat{\imath}_{\rm LL}=\argmax_{i\in\mathcal{C}}\sum_{j}[w_{ij}\ln p_j+(1-w_{ij})\ln(1-p_j)]$ over the full column pool, which is fixed for all subsets, and define the empirical excess error $\Delta_{\rm real}(d)\triangleq\hat{P}_{\rm err}^{\rm sum}(d)-P_{\rm floor}$.
$P_{\rm floor}$ is the error rate of a classifier and therefore at least the Bayes error rate, so $\Delta_{\rm real}$ underestimates the true excess error rate; it is a reference on the given binary classifiers, not the Bayes error rate.
For comparison, in a separate CIFAR-100 experiment ($M\in\{4,6,8,10,12,16\}$, $25$ seeds) a multi-class logistic regression trained on the same features has a higher error rate than LL decoding of the 2vR classifiers in $136$ of $150$ conditions and an equal one in $8$ (mean difference $0.011$; Appendix~\ref{app:exp_details}).
On the same classifier set, sum decoding is within $\pm0.005$ of LL decoding for $M=6$--$16$ (TOST equivalence test at level $0.05$) and identical to it for $M=4$, so the decoding rule is not the limiting factor (Appendix~\ref{app:exp_details}).

\subsection{Results}\label{subsec:exp_emdr_vs_d}

\begin{figure}[htbp]
  \centering
  \includegraphics[width=\linewidth]{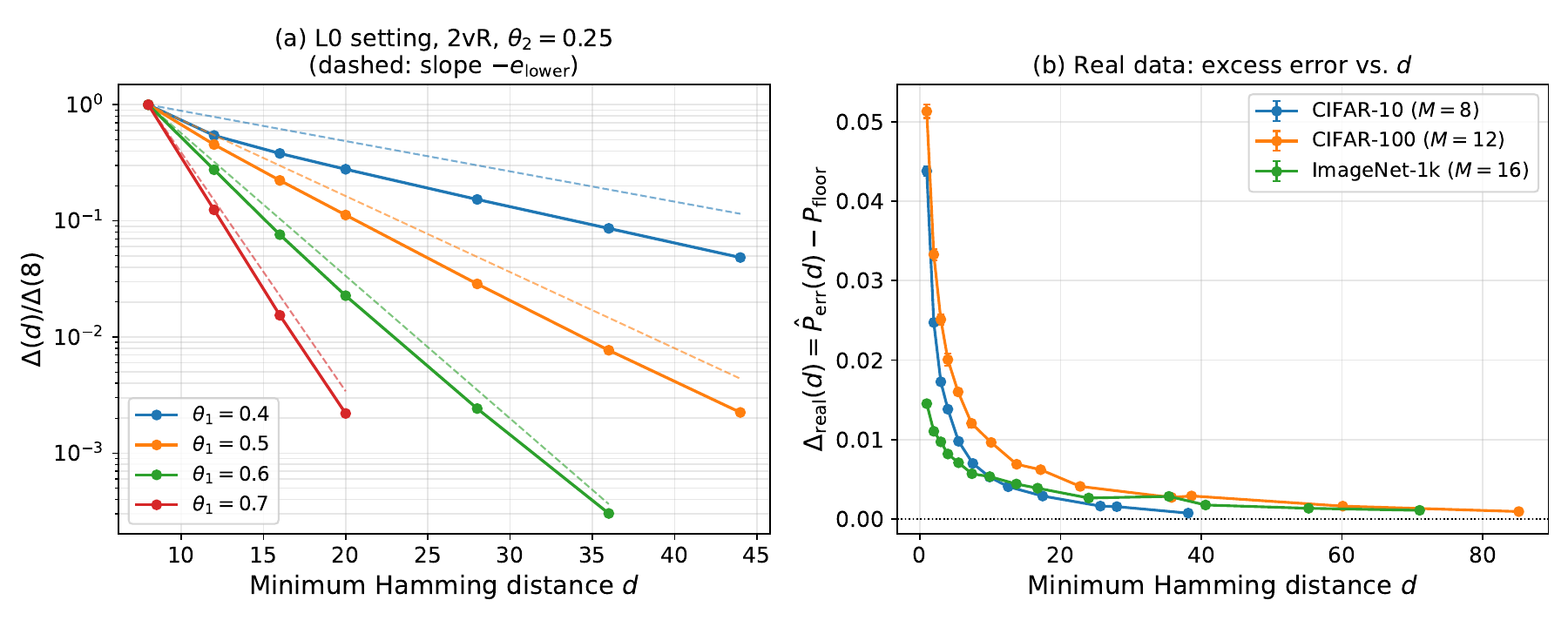}
  \caption{Excess error rate versus minimum Hamming distance $d$.
    (a)~L0 setting (2vR, $\theta_2=0.25$): $\Delta(d)/\Delta(8)$ on a log scale; the dashed lines have slope $-e_{\rm lower}$.
    (b)~Real data (ResNet-50 features, $S=40$ seeds): $\Delta_{\rm real}(d)=\hat{P}_{\rm err}^{\rm sum}(d)-P_{\rm floor}$, binned by the minimum distance of random column subsets (error bars: standard errors). These subsets are not equidistant, so their minimum distance is not a complete summary of the code (Remark~\ref{rem:spectrum}).}
  \label{fig:emdr_vs_d}
\end{figure}

\textbf{L0 setting (Figure \ref{fig:emdr_vs_d}(a)).}
$\Delta$ decays exponentially in $d$, faster than the slope $-e_{\rm lower}$ (the fitted slopes are $1.02$--$1.33$ times $e_{\rm lower}$, the most for the smallest $\theta_1$), as expected from the finite-length exponents of Appendix~\ref{app:numerical}, which approach $e_{\rm lower}$ from above for the 2vR code because of the subexponential prefactor of $P(E|\bm{\theta})$ (Appendix~\ref{subsec:app_main}).

\textbf{Real data (Figure \ref{fig:emdr_vs_d}(b)).}
For all three datasets $\Delta_{\rm real}(d)$ decreases as $d$ increases:
from $0.044$ ($d=1$) to $0.0007$ ($d\approx38$) for CIFAR-10 ($M=8$),
from $0.051$ to $0.0009$ ($d\approx85$) for CIFAR-100 ($M=12$),
and from $0.015$ to $0.0011$ ($d\approx71$) for ImageNet-1k ($M=16$).
The binned curves decrease monotonically except for one reversal of about $0.0002$ between adjacent bins on each of CIFAR-100 and ImageNet-1k, an order of magnitude smaller than the within-bin standard deviations ($\approx0.003$).
We define the saturation distance $d^*$ operationally as the smallest bin at which the mean of $\Delta_{\rm real}$ falls within one within-bin standard deviation of zero; this gives $d^*\approx38$, $36$ and $24$ for CIFAR-10, CIFAR-100 and ImageNet-1k. This is a practical threshold: beyond $d^*$ the remaining mean excess error is smaller than the spread across seeds and column subsets; it does not mean that $\Delta_{\rm real}$ is statistically indistinguishable from zero (the error bars in Figure~\ref{fig:emdr_vs_d}(b) are standard errors).
For CIFAR-10 ($M=8$) this criterion is met only near the largest distance probed, so the leveling-off is less pronounced there than for CIFAR-100 and ImageNet-1k.
Beyond $d^*$, increasing $d$ changes the error rate by less than the variation across seeds and column subsets while the number of classifiers keeps growing.
ImageNet-1k starts from a much smaller excess error ($0.015$ at $d=1$, against a reference error rate of $0.017$) and reaches $d^*$ at a smaller distance; the ResNet-50 features were trained on ImageNet-1k labels and therefore separate these classes well.
For $M=16$ the 2vR code has $d=28$, just beyond $d^*$ by this criterion, although the mean excess error there is still about $0.003$.

\textbf{Larger distance does not always help.}
On CIFAR-100 ($M\in\{8,10,12\}$, $25$ seeds), the 3vR code, whose distance $d(3,M)$ is $2.5$--$4.5$ times that of 2vR, has a slightly \emph{higher} error rate than 2vR ($+0.0020$, Wilcoxon signed-rank $p=1\times10^{-5}$, $75$ pairs; Appendix~\ref{app:exp_details}).
The bounds of Sections~\ref{sec:upper}--\ref{sec:tight} decrease with the distance and do not account for this small but systematic deficit, which Section~\ref{sec:discussion} examines through the effective distance of Section~\ref{subsec:correlated}.

\textbf{Scaling in $M$.}
Table~\ref{tab:imagenet_pe} increases the number of ImageNet-1k classes to $M=256$ (a separate run: PCA to $128$ dimensions, $S=3$ seeds, the 2vR and logM codes themselves rather than column subsets).
The logM code ($d=1$) degrades rapidly towards $\hat{P}_{\rm err}\to1$, whereas 2vR ($d=2(M-2)$) stays low, as the necessity and sufficiency parts of Theorem~\ref{th:main} lead one to expect; 2vR is also comparable to ($M=16$) or below ($M=128,256$) a multi-class logistic regression.
The LL-decoding reference was not computed in this run, so the table does not measure the distance to $P_{\rm floor}$, and its $M=16$ values are not comparable to Figure~\ref{fig:emdr_vs_d}(b), which uses a different training-set size and column subsets.

\begin{table}[tb]
  \centering
  \caption{ImageNet-1k: $\hat{P}_{\rm err}$ as the number of classes $M$ increases (ResNet-50 features, PCA-128, mean $\pm$ standard deviation over $S=3$ seeds). The multi-class logistic regression (softmax) is shown for reference.}
  \label{tab:imagenet_pe}
  \footnotesize
  \begin{tabular}{r r@{\;\;}l@{\;\;}l@{\;\;}l}
    \hline
    $M$ & $d_{\rm 2vR}$ & 2vR & logM ($d{=}1$) & multi-class LR (ref.) \\
    \hline
     16 &   28 & $0.037{\pm}0.017$ & $0.126{\pm}0.014$ & $0.037{\pm}0.017$ \\
    128 &  252 & $0.102{\pm}0.003$ & $0.682{\pm}0.018$ & $0.121{\pm}0.005$ \\
    256 &  508 & $0.128{\pm}0.007$ & $0.871{\pm}0.009$ & $0.156{\pm}0.008$ \\
    \hline
  \end{tabular}
\end{table}

\section{Discussion: What the Model Does and Does Not Capture}\label{sec:discussion}

The theory predicts the qualitative behavior on real data --- the decay with the minimum distance, the failure of short codes, and the leveling-off --- but not its quantitative form.
Four factors account for the gap, and each is visible in our results; the third is partly covered by Theorem~\ref{th:correlated}.

\emph{(i)~Spread of the posterior.}
Theorem~\ref{th:main} holds at each $\bm{\theta}$, whereas $\hat{P}_{\rm err}$ averages over the inputs.
At fixed $\bm{\theta}$ the pairwise term $Q(\rho_{1k})$ carries a prefactor of order $d^{-1/2}$ in addition to $e^{-dE_{\rm ECOC}(\bm{\theta},\sigma)}$ (Lemma~\ref{lem:Q_bounds}).
Averaging over inputs whose exponents $E_{\rm ECOC}(\bm{\theta},\sigma)$ approach their infimum adds a further algebraic factor, so the aggregate curve is not a single exponential.
Its decay is governed by the inputs with the smallest exponent, those close to the decision boundary.
This mirrors the role of the margin condition for the excess risk of plug-in classifiers \citep{audibert2007fast}.
On our curves a single exponential in $d$ fits poorly (a linear fit of $\ln\Delta_{\rm real}$ against $d$ over the bins of Figure~\ref{fig:emdr_vs_d}(b) gives $R^2=0.90$, $0.83$ and $0.89$ for CIFAR-10, CIFAR-100 and ImageNet-1k), and the rate cannot be read off them.

\emph{(ii)~Distance spectrum.}
The column subsets of Figure~\ref{fig:emdr_vs_d}(b) are not equidistant.
For such codes the error rate depends on the whole spectrum $\{A_\delta\}$ (Remark~\ref{rem:spectrum}), and the minimum distance is an incomplete summary.

\emph{(iii)~Correlated noise and the effective distance.}
By Theorem~\ref{th:correlated}, the quantity that governs the excess error rate under correlated noise is not $d_H$ but the effective distance $d_{\rm eff}(c_1,c_k)=d_H(c_1,c_k)^2/(\bm{s}_k^\top\Sigma\bm{s}_k)$, which depends on the code through the set of differing positions.
The class-level model of Theorem~\ref{th:correlated} accounts for the leveling-off, since $d_{\rm eff}$ saturates as $d_H$ grows, but not by itself for a deficit, which requires the effective distance of the longer code to be smaller for the pairs that matter.
To check this for the deficit of 3vR in Section~\ref{subsec:exp_emdr_vs_d}, we repeated the CIFAR-100 comparison with labels resampled from a calibrated teacher, so that the true posterior, and hence the log-odds residuals of each binary classifier under the logit-noise model of \citet{abe2026calibration}, are known; the classes, images and classifiers are as in Section~\ref{subsec:exp_setup} (Appendix~\ref{app:exp_details}).
The deficit persists in this setting ($+0.0012$ in excess error under LL decoding, positive in $57$ of $75$ seed-matched pairs, Wilcoxon $p=5\times10^{-7}$).
The residual covariance $\hat\Sigma$, estimated from held-out residuals, is largely explained by class-level noise: the class-level model of Theorem~\ref{th:correlated} fits it with $R^2=0.64$--$0.83$, and its saturation level $2/\sigma_{\rm c}^2$ is close to the measured $d_{\rm eff}$, so both codes are already in the saturated regime.
Accordingly, the effective distance of 3vR, averaged over pairs, is at most $1.04$ times that of 2vR for $M=8,10,12$, and its minimum over pairs is only $0.89$--$0.94$ times that of 2vR (medians over seeds), although its Hamming distance is $2.5$--$4.5$ times larger.
A Gaussian model with $\hat\Sigma$ predicts a higher error for 3vR in every seed, whereas Assumption~\ref{ass:eps} with a fitted $\sigma$ predicts a lower one; with a less confident teacher, for which the ratio of effective distances rises to $1.22$--$1.25$, 3vR becomes slightly better ($-0.0004$), as the Gaussian model predicts ($-0.0003$).
The predicted deficit in the first setting ($+0.0074$) is, however, about six times the observed one, and resampling the empirical residuals instead of using a Gaussian reverses its sign.
The effective distance is thus consistent with the absence of a gain from the larger Hamming distance, but it does not determine the size of the deficit, and with resampled residuals not even its sign; the residuals depend on the input (Appendix~\ref{app:exp_details}), which none of these predictors models.

\emph{(iv)~Reference decoder.}
$P_{\rm floor}$ is the error rate of LL decoding on the given classifiers, not the Bayes error rate; estimating the noise level of trained classifiers requires the true posterior, and no procedure for estimating it stably from their outputs alone has been established \citep{abe2026calibration}, so $e_{\rm lower}$ and $d^*$ cannot be computed from the theory on real data. Moreover, the theory concerns the product decoder \eqref{eq:Passerini}, whereas Figure~\ref{fig:emdr_vs_d}(b) uses sum decoding; their agreement within $\pm0.005$ (Section~\ref{subsec:exp_setup}) is empirical.

\textbf{Implications.}
Within these limitations, the theory gives a design rule: the minimum distance must grow so that $\kappa\,d_{\min}$ exceeds $\ln(M-1)$, and beyond the saturation distance a longer code only adds classifiers.
Since $e_{\rm lower}\propto1/\sigma^2$ and $\sigma$ depends on how the binary classifiers are trained, the saturation distance is not a property of the task alone.
The necessity part offers a possible explanation for why output layers that encode $M$ classes with $\lceil\log_2M\rceil$ bits lose accuracy unless redundancy is added \citep{oda2017neural}: with $d_{\min}=1$ the excess error rate does not vanish however the binary classifiers are improved, as long as their noise stays bounded away from zero.

\section{Conclusion}\label{sec:conclusion}

Extending the analysis of noisy binary classifiers by \citet{kumoi2023IJNS} from the pairwise error probability to the classification error rate, we characterized the excess error rate of ECOC over the MAP classifier by the minimum Hamming distance.
A bounded distance between the code words of the two most probable categories keeps the excess error rate away from zero uniformly in $M$; a minimum distance that outgrows $\ln M$ drives it to zero exponentially (for general codes under the condition of Proposition~\ref{prop:upper_general}); and for $n$vR codes ($n\geq2$) the exact exponent is the reliability function $E_{\rm ECOC}$, whose worst case over $\Omega(\lambda)$ is $\kappa_2(\lambda,\sigma)$.
Under correlated Gaussian noise the same statements hold with the effective distance in place of the Hamming distance, and class-level noise bounds the effective distance (Theorem~\ref{th:correlated}).
Future work includes the exponent of the Hadamard code and of other non-$n$vR codes, a bound for general codes in terms of the distance spectrum, a finite-length analysis including the prefactor, and the estimation of the effective distance of trained classifiers.

\ifdefined\ARXIVVERSION
\subsubsection*{Acknowledgments}
This work was supported by JSPS KAKENHI Grant Number JP24K20833.
\fi

\bibliographystyle{tmlr}
\bibliography{ref}

\appendix

\section{Proof of Lemma \ref{lem:eta_upper} (upper bound on $\eta$)}\label{app:eta_upper}

From the pair condition $w_{lj}=w_{lj'}$ ($l\neq 1,k$), the sum on the $f_{j'}$ side satisfies $\alpha_{j'}=1-\theta_1-\theta_k-\alpha_j$.
Substituting this and expanding $\eta_j$ as a function of $\alpha_j$ alone gives
\begin{align}\label{eq:eta_explicit}
\eta_j(\alpha_j;\theta_1,\theta_k)=\frac{(1-\theta_1-\alpha_j)(\theta_k+\alpha_j)}{(\theta_1+\alpha_j)(1-\theta_k-\alpha_j)},\qquad \alpha_j\in[0,1-\theta_1-\theta_k].
\end{align}
Differentiating its logarithm and combining the terms pairwise,
\begin{align}\label{eq:eta_logderiv}
\frac{\partial\ln\eta_j}{\partial\alpha_j}
=\frac{(\theta_1-\theta_k)(1-\theta_1-\theta_k-2\alpha_j)}{(\theta_1+\alpha_j)(\theta_k+\alpha_j)(1-\theta_1-\alpha_j)(1-\theta_k-\alpha_j)}.
\end{align}
The denominator is positive on the domain, so for $\theta_1>\theta_k$ the derivative is positive for $\alpha_j<(1-\theta_1-\theta_k)/2$ and negative for $\alpha_j>(1-\theta_1-\theta_k)/2$.
Hence $\eta_j$ is unimodal and attains its maximum at $\alpha_j=(1-\theta_1-\theta_k)/2$; substituting this value yields \eqref{eq:eta_upper}.
\hfill$\square$

\section{Proof of Lemma \ref{lem:eta_lower} (lower bound on $\eta$)}\label{app:eta_lower}

By \eqref{eq:eta_logderiv} (Appendix~\ref{app:eta_upper}), for $\theta_1>\theta_k$ the function $\eta_j$ in \eqref{eq:eta_explicit} increases on $[0,(1-\theta_1-\theta_k)/2]$ and decreases on $[(1-\theta_1-\theta_k)/2,1-\theta_1-\theta_k]$.
Hence its minimum on the closed interval is attained at an endpoint. We compute the values at the two endpoints.

At $\alpha_j=0$:
\begin{align}\label{eq:eta_at_0}
    \eta_j(0;\theta_1,\theta_k) = \frac{(1-\theta_1)\theta_k}{\theta_1(1-\theta_k)}.
\end{align}

At $\alpha_j=1-\theta_1-\theta_k$:
\begin{align}\label{eq:eta_at_max}
    \eta_j(1-\theta_1-\theta_k;\theta_1,\theta_k)
    = \frac{\theta_k(1-\theta_1)}{(1-\theta_k)\theta_1}.
\end{align}
This coincides with \eqref{eq:eta_at_0}.
Since $\eta_j$ takes the same value at the two endpoints of the domain, the minimum is this endpoint value:
\begin{align}
    \eta_j(\alpha_j;\theta_1,\theta_k) \geq \frac{(1-\theta_1)\theta_k}{\theta_1(1-\theta_k)}.
\end{align}
\hfill$\square$

\section{Proof of Lemma \ref{lem:c2_elower} (relation between the uniform and pointwise exponents)}\label{app:c2_elower}

Let $h(\bm{\theta})\triangleq\frac{\theta_1(1-\theta_2)}{(1-\theta_1)\theta_2}$, so that $E_{\rm ECOC}=\frac{1}{8\sigma^2}\ln^2 h$, and $\ln h>0$ since $\theta_1>\theta_2$.
Changing variables to $s=\theta_1+\theta_2\leq 1$ and $\delta=\theta_1-\theta_2\geq\lambda$,
\begin{align}
    \frac{\partial \ln h}{\partial s}
    = 2\delta\left[\frac{1}{(2-s)^2-\delta^2}-\frac{1}{s^2-\delta^2}\right]\leq 0,
\end{align}
where the inequality follows from $s\leq 1\leq 2-s$.
Hence $\ln h$ is monotone decreasing in $s$ and attains its minimum $\ln h=2\ln\frac{1+\delta}{1-\delta}$ at $s=1$.
Furthermore, $\ln\frac{1+\delta}{1-\delta}$ is monotone increasing in $\delta$, so it is minimized at $\delta=\lambda$, and
\begin{align}
    E_{\rm ECOC}(\bm{\theta},\sigma)
    &\geq \frac{1}{8\sigma^2}\left(2\ln\frac{1+\lambda}{1-\lambda}\right)^2 \nonumber\\
    &= \frac{1}{2\sigma^2}\ln^2\!\left(\frac{1+\lambda}{1-\lambda}\right)
    = \kappa_2(\lambda,\sigma).
\end{align}
\hfill$\square$

\section{Proof of Corollary \ref{th:exhaustive_tight} (asymptotic evaluation of the Exhaustive Code)}\label{app:exhaustive}

\begin{corollary}[Asymptotic evaluation of the Exhaustive Code]\label{th:exhaustive_tight}
For $\bm{\theta}\in\Omega(\lambda)$, for any $\zeta>0$ and sufficiently large $M$:

Upper bound:
\begin{align}\label{eq:exh_upper}
    \Delta(\bm{\theta})
    \leq \exp\!\left\{-N_{\rm exh}\left(\frac{\kappa_2(\lambda,\sigma)}{2}-\zeta\right)\right\}.
\end{align}

Lower bound:
\begin{align}\label{eq:exh_lower}
    \Delta(\bm{\theta})
    \geq \Theta\!\left(\frac{1}{\sqrt{N_{\rm exh}}}\right)
    \exp\!\left\{-N_{\rm exh}\,\delta_{\rm exh}\,e_{\rm exh}(\bm{\theta})\right\},
\end{align}
where $N_{\rm exh}=2^{M-1}-1$, $\delta_{\rm exh}=d_{\rm exh}/N_{\rm exh}\in(1/2,1]$, and $e_{\rm exh}(\bm{\theta})\triangleq\frac{1}{8\sigma^2}\ln^2\frac{\theta_1(1-\theta_2)}{(1-\theta_1)\theta_2}$.
\end{corollary}

\textit{Upper bound.} Applying Proposition \ref{pro:mu_upper_nvR} to each block and summing, $\mu_{1k}\leq d_{\rm exh}\ln\frac{1-\theta_1+\theta_k}{1+\theta_1-\theta_k}$, and as in Corollary \ref{cor:upper_nvR} we obtain $\Delta(\bm{\theta})\leq(M-1)\exp(-d_{\rm exh}\,\kappa_2(\lambda,\sigma))$.
From $d_{\rm exh}=N_{\rm exh}\delta_{\rm exh}$, $\delta_{\rm exh}>1/2$, and $M-1=\log_2(N_{\rm exh}+1)\leq e^{\zeta N_{\rm exh}}$ (for sufficiently large $M$), we obtain \eqref{eq:exh_upper}.

\textit{Lower bound.} Applying Proposition \ref{pro:mu_lower_nvR} to each block (for the 1vR block it holds with equality, Section~\ref{sec:specific}) and summing, $\mu_{1k}\geq \frac{d_{\rm exh}}{2}\ln\frac{(1-\theta_1)\theta_k}{\theta_1(1-\theta_k)}$, and by the same argument as Proposition \ref{th:EMDR_lower} we obtain $\Delta(\bm{\theta})\geq\frac{\lambda}{2\sqrt{2\pi}\,\rho^{\rm U}}\exp(-(\rho^{\rm U})^2/2)$ with $(\rho^{\rm U})^2/2=d_{\rm exh}\,e_{\rm exh}(\bm{\theta})$.
Since $\rho^{\rm U}=\sqrt{2d_{\rm exh}\,e_{\rm exh}(\bm{\theta})}=\Theta(\sqrt{N_{\rm exh}})$, the prefactor is $\Theta(1/\sqrt{N_{\rm exh}})$.
\hfill$\square$

\section{Proof of Lemma \ref{lem:mu_asymptotic} (asymptotically exact value of $\mu_{12}$)}\label{app:mu_asymptotic}

\textit{Step 1: exact expression for $\mu_{12}$.}

The positive-side group of a classifier $f_j$ separating $c_1$ and $c_2$ has the form $\{c_1\}\cup\mathcal{S}$ ($|\mathcal{S}|=n-1$, $\mathcal{S}\subseteq\{3,\ldots,M\}$).
Setting $\alpha_{\mathcal{S}}\triangleq\sum_{i\in\mathcal{S}}\theta_i$,
\begin{align}\label{eq:mu_exact}
    \mu_{12} = \sum_{\mathcal{S}\subseteq\{3,\ldots,M\},|\mathcal{S}|=n-1}\ln\frac{(1-\theta_1-\alpha_{\mathcal{S}})(\theta_2+\alpha_{\mathcal{S}})}{(\theta_1+\alpha_{\mathcal{S}})(1-\theta_2-\alpha_{\mathcal{S}})}.
\end{align}
This expression assumes $M>2n$. For $n=M/2$ each pair appears in the sum twice, as $\mathcal{S}$ and as its complement in $\{3,\ldots,M\}$, with equal terms, so the right-hand side must be multiplied by $1/2$; this is the value used for $M=4$ in Table~\ref{tab:app_lemma_verification}.
In both cases the number of distinct terms is $d(n,M)/2$, so $\mu_{12}/(d(n,M)/2)$ is the average of $\gamma(\alpha_{\mathcal{S}})$ over all sets $\mathcal{S}\subseteq\{3,\ldots,M\}$ with $|\mathcal{S}|=n-1$, where
\begin{equation}\label{eq:gamma_def}
\gamma(\alpha)\triangleq\ln\frac{(1-\theta_1-\alpha)(\theta_2+\alpha)}{(\theta_1+\alpha)(1-\theta_2-\alpha)},\qquad\alpha\in[0,1-\theta_1-\theta_2].
\end{equation}

\textit{Step 2: two-sided bound on $\gamma$.}

Since $\gamma=\ln\eta_j$ with $\eta_j$ as in \eqref{eq:eta_explicit} ($\theta_k=\theta_2$), Lemma~\ref{lem:eta_lower} gives $\gamma(\alpha)\geq\gamma(0)$ on the whole domain; by \eqref{eq:eta_logderiv}, $\gamma$ is strictly increasing and then strictly decreasing between two endpoints at which it takes the same value, so $\gamma(\alpha)>\gamma(0)$ for $0<\alpha<1-\theta_1-\theta_2$.
For the upper bound, write $\gamma(\alpha)-\gamma(0)$ as the sum of
$\ln\frac{1-\theta_1-\alpha}{1-\theta_1}\leq-\frac{\alpha}{1-\theta_1}$,
$\ln\frac{\theta_2+\alpha}{\theta_2}\leq\frac{\alpha}{\theta_2}$,
$-\ln\frac{\theta_1+\alpha}{\theta_1}\leq-\frac{\alpha}{\theta_1+\alpha}\leq-\frac{\alpha}{1-\theta_2}$ and
$\ln\frac{1-\theta_2}{1-\theta_2-\alpha}\leq\frac{\alpha}{1-\theta_2-\alpha}\leq\frac{\alpha}{\theta_1}$,
where we used $\ln x\leq x-1$ (and, for the third, $\ln x\geq1-1/x$) together with $\theta_1+\alpha\leq1-\theta_2$ and $1-\theta_2-\alpha\geq\theta_1$. Hence, with $c'\triangleq\frac{1}{\theta_1}+\frac{1}{\theta_2}-\frac{1}{1-\theta_1}-\frac{1}{1-\theta_2}\geq0$,
\begin{equation}\label{eq:gamma_two_sided}
    0\leq\gamma(\alpha)-\gamma(0)\leq c'\alpha,\qquad\alpha\in[0,1-\theta_1-\theta_2].
\end{equation}

\textit{Step 3: evaluation of $\bar{\alpha}_{\mathcal{S}}$.}

Each $i\in\{3,\ldots,M\}$ is contained in $\binom{M-3}{n-2}$ of the $\binom{M-2}{n-1}$ sets $\mathcal{S}$, so
\begin{align}\label{eq:alpha_bar}
    \bar{\alpha}_{\mathcal{S}} = \frac{n-1}{M-2}(1-\theta_1-\theta_2).
\end{align}
This is an identity that uses only $\sum_{i\geq3}\theta_i=1-\theta_1-\theta_2$.

\textit{Step 4: derivation of the claim.}

Define the error from the asymptotic target value at finite $M$ by
\begin{equation}\label{eq:rM_def}
r(M)\triangleq\frac{\mu_{12}}{d(n,M)/2}-\ln\frac{(1-\theta_1)\theta_2}{\theta_1(1-\theta_2)}=\overline{\gamma(\alpha_{\mathcal{S}})-\gamma(0)}.
\end{equation}
Averaging \eqref{eq:gamma_two_sided} over $\mathcal{S}$ and using \eqref{eq:alpha_bar},
\begin{equation}
    0\leq r(M)\leq c'\,\bar{\alpha}_{\mathcal{S}}=\frac{\beta(\theta_1,\theta_2,n)}{M-2}
\end{equation}
for every $M\geq2n$, which is \eqref{eq:mu_asymptotic_rate} and implies \eqref{eq:mu_asymptotic}.
If $\theta_i>0$ for all $i\geq3$, then $0<\alpha_{\mathcal{S}}<1-\theta_1-\theta_2$ for every $\mathcal{S}$ (since $n-1<M-2$), so every term is strictly positive by Step~2 and $r(M)>0$ for every $M\geq2n$; that is, at finite $M$, $\mu_{12}/(d/2)$ lies on the larger side of the asymptotic value (smaller in absolute value).
Consequently $\rho_{12}<\rho^{\rm U}_{12}$ and $\rho_{12}^2/(2d(n,M))<e_{\rm lower}$ at every finite $M$: the correction lowers the Gaussian exponent of the pairwise error.

The leading term of $r(M)$ is given by a Taylor expansion around $\alpha=0$. By \eqref{eq:eta_logderiv},
\begin{align}
    \gamma'(0) = -\frac{1}{1-\theta_1}+\frac{1}{\theta_2}-\frac{1}{\theta_1}+\frac{1}{1-\theta_2}
    =\frac{(\theta_1-\theta_2)(1-\theta_1-\theta_2)}{\theta_1\theta_2(1-\theta_1)(1-\theta_2)},
\end{align}
which is positive if and only if $1-\theta_1-\theta_2>0$ (given $\theta_1>\theta_2$); this holds in the setting of the lemma.
Under $\theta_i=O(1/M)$ ($i\geq3$) we have $\overline{\alpha_{\mathcal{S}}^2}=O(1/M^2)$, so
\begin{align}
    r(M)= \frac{\gamma'(0)(n-1)(1-\theta_1-\theta_2)}{M-2} + O\!\left(\frac{1}{M^2}\right),
\end{align}
and $r(M)\times M$ converges to the constant $\gamma'(0)(n-1)(1-\theta_1-\theta_2)$, which is at most $\beta(\theta_1,\theta_2,n)$ since $c'-\gamma'(0)=2/\theta_1-2/(1-\theta_2)\geq0$; this can be confirmed numerically (Appendix \ref{subsec:app_lemma}).

\textit{Step 5: contribution to the reliability function.}

Since $r(M)\to 0$, $\rho_{12}^2/(2d(n,M))\to\frac{1}{8\sigma^2}\ln^2\frac{\theta_1(1-\theta_2)}{(1-\theta_1)\theta_2}=e_{\rm lower}$.
\hfill$\square$

\section{Proof of Proposition \ref{prop:reliability} (reliability function)}\label{app:reliability}

Let $e_{\rm lower}\triangleq\frac{1}{8\sigma^2}\ln^2\!\left(\frac{\theta_1(1-\theta_2)}{(1-\theta_1)\theta_2}\right)$.

\textit{The competitors $k\geq3$ are negligible.}
Let $d=d(n,M)$, and let $c_0$ be a constant with $\theta_i\leq c_0/M$ for $i\geq3$ (Definition~\ref{def:reliability}).
For $M>2n$, as in \eqref{eq:mu_exact} with $k$ in place of $2$, $\mu_{1k}=\sum_{\mathcal{S}}\gamma_k(\alpha_{\mathcal{S}})$, where $\mathcal{S}$ ranges over the $d/2$ subsets of $\{2,\ldots,M\}\setminus\{k\}$ of size $n-1$, $\gamma_k(\alpha)\triangleq\ln\frac{(1-\theta_1-\alpha)(\theta_k+\alpha)}{(\theta_1+\alpha)(1-\theta_k-\alpha)}$, and every term is negative by Lemma~\ref{lem:eta_upper}.
If $\theta_k+\alpha_{\mathcal{S}}\leq1/2$, then $-\gamma_k(\alpha_{\mathcal{S}})\geq\ln\frac{\theta_1}{2(\theta_k+\alpha_{\mathcal{S}})}$.
If $2\notin\mathcal{S}$, then $\theta_k+\alpha_{\mathcal{S}}\leq nc_0/M$, and the fraction of the sets $\mathcal{S}$ containing $2$ is $(n-1)/(M-2)$.
Hence
\begin{equation}\label{eq:kge3}
    \min_{k\geq3}\frac{-\mu_{1k}}{d}\ \geq\ \frac12\Bigl(1-\frac{n-1}{M-2}\Bigr)\Bigl(\ln M-\ln\frac{2nc_0}{\theta_1}\Bigr)\ \longrightarrow\ \infty,
\end{equation}
which holds for every $M>2n$ (when the right-hand side is negative it is trivial, the left-hand side being nonnegative), whereas $-\mu_{12}/d$ converges by Lemma~\ref{lem:mu_asymptotic}; since $\rho_{1k}=-\mu_{1k}/(\sigma\sqrt{d})$, we have $\rho_{\min}=\rho_{12}$ for all sufficiently large $M$.

\textit{Upper and lower bounds.}
By the union bound and the Chernoff bound (the upper bound of Lemma \ref{lem:Q_bounds}), we obtain an \emph{upper} bound on $P(E|\bm{\theta})$,
\begin{align}\label{eq:app_union}
    P(E|\bm{\theta}) \leq (M-1)\exp\!\left(-\frac{\rho_{\min}^2}{2}\right),
\end{align}
which yields a \emph{lower} bound on $-\ln P(E|\bm{\theta})$.
We have $\rho_{\min}=\rho_{12}$ for all sufficiently large $M$, $\rho_{12}^2/(2d(n,M))\to e_{\rm lower}$ by Lemma \ref{lem:mu_asymptotic} (Appendix \ref{app:mu_asymptotic}), and $(M-1)\leq e^{\zeta d(n,M)}$ for any $\zeta>0$ (since $\ln(M-1)=o(d(n,M))$), so
\begin{align}
    \liminf_{M\to\infty}\frac{-\ln P(E|\bm{\theta})}{d(n,M)} \geq e_{\rm lower}.
\end{align}

Conversely, by $\{R_{12}\leq1\}\subseteq E$ and the lower bound of Lemma \ref{lem:Q_bounds} (when $\rho_{12}\geq1$), we obtain a \emph{lower} bound on $P(E|\bm{\theta})$,
\begin{align}
    P(E|\bm{\theta}) \geq \frac{1}{2\sqrt{2\pi}\rho_{12}}\exp\!\left(-\frac{\rho_{12}^2}{2}\right),
\end{align}
which yields an \emph{upper} bound on $-\ln P(E|\bm{\theta})$.
Since $\ln\rho_{12}=O(\ln d(n,M))$ is sublinear in $d(n,M)$,
\begin{align}
    \limsup_{M\to\infty}\frac{-\ln P(E|\bm{\theta})}{d(n,M)} \leq e_{\rm lower}.
\end{align}

Since $\liminf\leq\limsup$ always holds, the two bounds above force $\liminf=\limsup=e_{\rm lower}$, so the limit exists and $E_{\rm ECOC}(\bm{\theta},\sigma) = e_{\rm lower}$.
\hfill$\square$

\section{Proof of Theorem \ref{th:correlated} (correlated Gaussian noise)}\label{app:correlated}

\textit{Part 1, sufficiency and necessity.}
For a general covariance, \eqref{eq:g_comp} and \eqref{eq:channel_model} give $P(R_{1k}\leq1|\bm{\theta})=Q(\rho_{1k})$ with
\begin{equation}
    \rho_{1k}\triangleq\frac{-\mu_{1k}}{\sqrt{\bm{s}_k^\top\Sigma\bm{s}_k}}=\frac{-\mu_{1k}}{d_H(c_1,c_k)}\sqrt{d_{\rm eff}(c_1,c_k)},
\end{equation}
which replaces \eqref{eq:mu_rho}.
Proposition~\ref{prop:Delta_basic} involves the noise only through $P(E|\bm{\theta})$, and Propositions~\ref{pro:mu_upper_general}, \ref{pro:mu_upper_nvR} and \ref{pro:mu_lower_nvR} and the bound \eqref{eq:mu_col_bound} concern $\mu_{1k}$ alone.
By the union bound and Lemma~\ref{lem:Q_bounds}, $P(E|\bm{\theta})\leq\sum_{k\neq1}\exp(-\rho_{1k}^2/2)$.
By \eqref{eq:mu_upper_general}, $-\mu_{1k}/d_H(c_1,c_k)\geq\ln\frac{\theta_1+\omega(n_{\min})}{1-\theta_1-\omega(n_{\min})}$ under the conditions of Proposition~\ref{prop:upper_general}, and by \eqref{eq:mu_upper_nvR}, $-\mu_{1k}/d(n,M)\geq\ln\frac{1+\lambda}{1-\lambda}$ for $n$vR codes; in both cases $\rho_{1k}^2/2\geq\bar\kappa\,d_{\rm eff}(c_1,c_k)$, which gives the upper bound in \eqref{eq:correlated_bounds}.
For the lower bound, $\{R_{12}\leq1\}\subseteq E$ gives $P(E|\bm{\theta})\geq Q(\rho_{12})$, and \eqref{eq:mu_col_bound} gives $-\mu_{12}\leq d_H(c_1,c_2)B_2(\bm{\theta})$, so $\rho_{12}\leq\sqrt{d_{\rm eff}(c_1,c_2)}\,B_2(\bm{\theta})$; the claim follows from Proposition~\ref{prop:Delta_basic} since $Q$ is decreasing.

\textit{Part 1, tightness.}
Let $d=d(n,M)$ with $n\geq2$ fixed, take the family of Definition~\ref{def:reliability}, and let $V$ bound $\sigma^2_{\rm eff}(1,k)$.
By \eqref{eq:kge3} (Appendix~\ref{app:reliability}), which concerns $\mu_{1k}$ alone, $-\mu_{1k}/d\to\infty$ uniformly in $k\geq3$, so that $\rho_{1k}^2/(2d)\geq(\mu_{1k}/d)^2/(2V)\to\infty$ uniformly in $k\geq3$.
By Lemma~\ref{lem:mu_asymptotic}, $\rho_{12}^2/(2d)=(\mu_{12}/d)^2/(2\sigma^2_{\rm eff}(1,2))\to E_{\rm ECOC}(\bm{\theta},\sqrt{v})$.
Therefore $\rho_{\min}=\rho_{12}$ for all sufficiently large $M$, and the argument of Appendix~\ref{app:reliability} applies with $\sigma\sqrt{d}$ replaced by $(\bm{s}_2^\top\Sigma\bm{s}_2)^{1/2}$: $Q(\rho_{12})\leq P(E|\bm{\theta})\leq(M-1)\exp(-\rho_{12}^2/2)$ and $\ln(M-1)=o(d)$ give $-\ln P(E|\bm{\theta})/d\to E_{\rm ECOC}(\bm{\theta},\sqrt{v})$, and Proposition~\ref{prop:Delta_basic} transfers the limit to $\Delta(\bm{\theta})$ as in Corollary~\ref{cor:EMDR_reliability}.

\textit{Part 2.}
Since $\mathrm{Cov}(\varepsilon_j,\varepsilon_{j'})=\sigma_0^2\mathbf{1}_{\{j=j'\}}+\sigma_{\rm c}^2\sum_iw_{ij}w_{ij'}+\sigma_{\rm z}^2$, we have $\Sigma=\sigma_0^2I+\sigma_{\rm c}^2W^\top W+\sigma_{\rm z}^2\mathbf{1}\mathbf{1}^\top$.
With $\bm{e}_i$ the $i$-th unit vector of $\mathbb{R}^M$, $d_H(c_1,c_k)=\lVert\bm{s}_k\rVert^2=(\bm{e}_k-\bm{e}_1)^\top W\bm{s}_k\leq\sqrt{2}\,\lVert W\bm{s}_k\rVert$ by the Cauchy--Schwarz inequality, so $\bm{s}_k^\top\Sigma\bm{s}_k\geq\sigma_{\rm c}^2\lVert W\bm{s}_k\rVert^2\geq\sigma_{\rm c}^2d_H(c_1,c_k)^2/2$, that is, $d_{\rm eff}(c_1,c_k)\leq2/\sigma_{\rm c}^2$; the lower bound on $\Delta(\bm{\theta})$ follows from part~1.
Let $W$ be equidistant with distance $d$ and row weight $\ell$.
Then $WW^\top=(d/2)I+(\ell-d/2)\mathbf{1}\mathbf{1}^\top$, and the noise part of $L_i$ in \eqref{eq:sufficient_stat} is
\begin{equation}
    \sum_jw_{ij}\varepsilon_j=\sum_jw_{ij}\xi_j+\frac{d}{2}\chi_i+\Bigl(\ell-\frac{d}{2}\Bigr)\sum_{l}\chi_l+\ell z .
\end{equation}
The last two terms are common to all $i$.
The covariance of the first two, $\sigma_0^2WW^\top+\sigma_{\rm c}^2(d/2)^2I$, differs from $(\sigma_0^2+\sigma_{\rm c}^2d/2)\,WW^\top$, the covariance under Assumption~\ref{ass:eps} with $\sigma^2=\sigma_0^2+\sigma_{\rm c}^2d/2$, by a multiple of $\mathbf{1}\mathbf{1}^\top$, which cancels in every difference $L_i-L_k$.
The score differences are jointly Gaussian with the same means and covariances in the two models, so $\hat{i}$ has the same distribution.
If $\sigma_{\rm c}=0$, only the term $\ell z$ remains, and equal row weights suffice.
\hfill$\square$

\section{Numerical Verification}\label{app:numerical}

In this appendix we present the numerical verification of the theoretical results up to Section \ref{sec:tight}.
The setting here corresponds to the L0 setting (Section \ref{subsec:exp_setup}) that exactly satisfies the assumptions of the theory (i.i.d. Gaussian noise of known variance $\sigma^2$ in the log-odds space), in which the value $e_{\rm lower}$ of the reliability function can be verified quantitatively.
We take $n=2$ (2vR code), $\theta_1=0.5$, $\theta_2=0.25$ (so $\lambda=\theta_1-\theta_2=0.25$), $\sigma=1.0$, and allocate the posterior probabilities of the remaining categories uniformly as $\theta_i=(1-\theta_1-\theta_2)/(M-2)$ ($i\geq 3$).
We used 500,000 trials for the Monte Carlo estimate for each $M$.

\subsection{Verification of the lemma: asymptotic behavior of $\mu_{12}/(d/2)$}\label{subsec:app_lemma}

Lemma \ref{lem:mu_asymptotic} claims that, as $M\to\infty$ ($n$ fixed),
\begin{equation}
  \frac{\mu_{12}}{d(n,M)/2}
  \;\longrightarrow\;
  \ln\frac{(1-\theta_1)\theta_2}{\theta_1(1-\theta_2)}
  \label{eq:app_lemma_limit}
\end{equation}
and guarantees that the error is $O(1/M)$.
Table \ref{tab:app_lemma_verification} compares the exact numerical value of the left-hand side with the asymptotic target value.

\begin{table}[tb]
  \centering
  \caption{Numerical verification of Lemma \ref{lem:mu_asymptotic}
    ($n=2$, $\theta_1=0.5$, $\theta_2=0.25$, $\sigma=1.0$).}
  \label{tab:app_lemma_verification}
  \begin{tabular}{rr@{\quad}c@{\quad}c@{\quad}c@{\quad}c}
    \hline
    $M$ & $d(2,M)$ & $\mu_{12}/(d/2)$ & asymptotic target & error $r(M)$ & $|r(M)|\times M$ \\
    \hline
     4 &   2 & $-1.02165$ & $-1.09861$ & $+0.07696$ & $0.308$ \\
     6 &   8 & $-1.03977$ & $-1.09861$ & $+0.05884$ & $0.353$ \\
    10 &  16 & $-1.06343$ & $-1.09861$ & $+0.03518$ & $0.352$ \\
    20 &  36 & $-1.08142$ & $-1.09861$ & $+0.01719$ & $0.344$ \\
    50 &  96 & $-1.09186$ & $-1.09861$ & $+0.00675$ & $0.338$ \\
   100 & 196 & $-1.09526$ & $-1.09861$ & $+0.00336$ & $0.336$ \\
   500 & 996 & $-1.09794$ & $-1.09861$ & $+0.00067$ & $0.334$ \\
    \hline
  \end{tabular}
\end{table}

$|r(M)|\times M$ converges to about $0.333$ ($=1/3$) over $M=4$ to $M=500$, confirming that the error is $O(1/M)$; every entry also satisfies the explicit bound $0<r(M)\leq\beta/(M-2)$ of the lemma, here with $\beta=2/3$.
Moreover, the error $r(M)$ is positive (for every $M$, by Lemma~\ref{lem:mu_asymptotic}), so $\mu_{12}/(d/2)$ at finite $M$ is larger than the asymptotic target, i.e., smaller in absolute value.
Hence $\rho_{12}<\rho^{\rm U}_{12}$ and $\rho_{12}^2/(2d(n,M))<e_{\rm lower}$ at every finite $M$: the correction of Lemma~\ref{lem:mu_asymptotic} lowers the Gaussian exponent of the pairwise error $Q(\rho_{12})$ below $e_{\rm lower}$ and raises $Q(\rho_{12})$ above $Q(\rho^{\rm U}_{12})$.
For example, at $M=20$, $\rho_{12}^2/(2d)=0.1462<e_{\rm lower}=0.1509$ and $Q(\rho_{12})=5.89\times10^{-4}>Q(\rho^{\rm U}_{12})=4.91\times10^{-4}$.

\subsection{Verification of the main theorem: convergence of the reliability function}\label{subsec:app_main}

Proposition \ref{prop:reliability} claims that, for an $n$vR code with $n\geq 2$ (here the 2vR code),
\begin{equation}
  \lim_{M\to\infty} \frac{-\ln P(E|\bm{\theta})}{d(n,M)}
  \;=\;
  e_{\rm lower}
  \;=\;
  \frac{1}{8\sigma^2}\ln^2\!\frac{\theta_1(1-\theta_2)}{(1-\theta_1)\theta_2}.
\end{equation}
In this setting $e_{\rm lower}=0.15087$, which is larger than the uniform worst-case exponent $\kappa_2(\lambda,\sigma)=0.13047$ over $\Omega(\lambda)$ (see Lemma \ref{lem:c2_elower}) by about $15.6\%$.

Table \ref{tab:app_main_theorem_verification} compares the Monte Carlo estimate of the exponent $-\ln P(E|\bm{\theta})_{\rm MC}/d$ with both formulas.

\begin{table}[tb]
  \centering
  \caption{Numerical verification of the reliability function
    ($n=2$, $\theta_1=0.5$, $\theta_2=0.25$, $\sigma=1.0$).}
  \label{tab:app_main_theorem_verification}
  \begin{tabular}{rr@{\quad}c@{\quad}c@{\quad}c@{\quad}c@{\quad}c}
    \hline
    $M$ & $d$ & $P(E)_{\rm MC}$ & $\frac{-\ln P(E)_{\rm MC}}{d}$ &
    $e_{\rm lower}$ & $\kappa_2$ (uniform) & $-\ln Q(\rho^U)/d$ \\
    \hline
     4 &   2 & $0.3384$ & $0.5418$ & $0.1509$ & $0.1305$ & $0.7602$ \\
     6 &   8 & $0.0728$ & $0.3276$ & $0.1509$ & $0.1305$ & $0.3514$ \\
     8 &  12 & $0.0341$ & $0.2816$ & $0.1509$ & $0.1305$ & $0.2964$ \\
    10 &  16 & $0.0166$ & $0.2563$ & $0.1509$ & $0.1305$ & $0.2668$ \\
    15 &  26 & $0.0030$ & $0.2239$ & $0.1509$ & $0.1305$ & $0.2297$ \\
    20 &  36 & $0.0006$ & $0.2061$ & $0.1509$ & $0.1305$ & $0.2117$ \\
    \hline
  \end{tabular}
  \vspace{2pt}
  \begin{minipage}{0.9\linewidth}
    \footnotesize
    $-\ln Q(\rho^U)/d$: the exponent of the lower bound $Q(\rho^{\rm U}_{12})$ of Lemma~\ref{th:PE_nvR_sandwich}. Since $(\rho^{\rm U}_{12})^2/2=d\,e_{\rm lower}$ exactly, it exceeds $e_{\rm lower}$ only through the prefactor of $Q$ and converges to $e_{\rm lower}$ as $d\to\infty$.
    At $M=4$ the 2vR code has $n=M/2$, so $d(2,4)=\binom{2}{1}=2$ by \eqref{eq:d_nvR}.
  \end{minipage}
\end{table}

From the table we obtain the following observations.
(1)~The Monte Carlo exponent decreases monotonically as $M$ increases and converges toward $e_{\rm lower}=0.15087$.
(2)~At finite $M$ the Monte Carlo exponent is larger than $e_{\rm lower}$.
This excess is not caused by the correction $r(M)$ of Lemma~\ref{lem:mu_asymptotic}, which acts in the opposite direction: $r(M)>0$ makes $|\mu_{12}|$, and hence $\rho_{12}$, smaller, so that $\rho_{12}^2/(2d)<e_{\rm lower}$ (previous subsection).
It is caused by the subexponential prefactor of $Q$.
By the Mills-ratio bounds $\phi(x)\,x/(x^2+1)\leq Q(x)\leq\phi(x)/x$ ($x>0$, $\phi$ the standard normal density),
$-\ln Q(\rho)/d=\rho^2/(2d)+\ln(\sqrt{2\pi}\,\rho)/d+O(1/(\rho^2d))$,
and for $n=2$ the term $\ln(\sqrt{2\pi}\,\rho_{12})/d=\Theta(\ln d/d)$ exceeds the deficit $e_{\rm lower}-\rho_{12}^2/(2d)=O(1/M)=O(1/d)$.
Indeed, $-\ln Q(\rho_{12})/d=0.3311$, $0.2557$ and $0.2066$ for $M=6$, $10$ and $20$, in close agreement, within Monte Carlo error, with the Monte Carlo exponents $0.3276$, $0.2563$ and $0.2061$.
(3)~The uniform exponent $\kappa_2=0.13047$ is below the Monte Carlo exponent at every $M$, consistent with the relation $E_{\rm ECOC}\geq \kappa_2$ of Lemma \ref{lem:c2_elower}.

\subsection{Verification of the consistency of the upper and lower bounds}\label{subsec:app_bounds}

We check the inequalities
\begin{equation}
  Q(\rho^{\rm U}_{12}) \;\leq\; P(E|\bm{\theta}) \;\leq\; (M-1)\,Q(\rho^{\rm L}).
  \label{eq:app_sandwich_bounds}
\end{equation}
The left inequality is the lower bound of Lemma \ref{th:PE_nvR_sandwich}.
The right inequality follows from the union bound and $\rho_{1k}\geq\rho^{\rm L}$ (Eq.~\eqref{eq:rho_L}); it is tighter than the Chernoff form $(M-1)\exp(-d\,\kappa_2)$ of Lemma \ref{th:PE_nvR_sandwich}, which is obtained from it by $Q(x)\leq e^{-x^2/2}$.
Table \ref{tab:app_bounds_consistency} compares the upper and lower bounds with the Monte Carlo estimates.

\begin{table}[tb]
  \centering
  \caption{Verification of the consistency of the upper and lower bounds
    ($n=2$, $\theta_1=0.5$, $\theta_2=0.25$, $\sigma=1.0$, 500,000 trials).}
  \label{tab:app_bounds_consistency}
  \begin{tabular}{rr@{\quad}c@{\quad}c@{\quad}c@{\quad}c}
    \hline
    $M$ & $d$ & $P(E)_{\rm MC}$ & lower $\,Q(\rho^U)$ & upper $(M{-}1)Q(\rho^L)$ & consistency \\
    \hline
     4 &   2 & $0.3384$ & $0.2186$ & $0.7051$ & \checkmark \\
     5 &   6 & $0.1165$ & $0.0892$ & $0.4217$ & \checkmark \\
     6 &   8 & $0.0728$ & $0.0601$ & $0.3713$ & \checkmark \\
     8 &  12 & $0.0341$ & $0.0285$ & $0.2688$ & \checkmark \\
    10 &  16 & $0.0166$ & $0.0140$ & $0.1846$ & \checkmark \\
    15 &  26 & $0.0030$ & $0.0025$ & $0.0644$ & \checkmark \\
    18 &  32 & $0.0011$ & $0.0009$ & $0.0328$ & \checkmark \\
    20 &  36 & $0.0006$ & $0.0005$ & $0.0207$ & \checkmark \\
    \hline
  \end{tabular}
\end{table}

For all verified $M$ ($4\leq M\leq 20$), the Monte Carlo estimate falls between the upper and lower bounds (\checkmark).

As $M$ increases, the absolute gap between the two bounds shrinks, but their ratio grows, from $3.2$ at $M=4$ to $42$ at $M=20$.
The normalized gap $(\ln[(M-1)Q(\rho^{\rm L})]-\ln Q(\rho^{\rm U}_{12}))/d$ decreases from $0.59$ to $0.10$ over the same range, and it tends to $e_{\rm lower}-\kappa_2(\lambda,\sigma)=0.0204>0$ rather than to zero.
Neither bound involves $r(M)$: the lower bound uses $\rho^{\rm U}_{12}$, i.e., the asymptotic value of $\mu_{12}$, and the upper bound uses the uniform $\rho^{\rm L}$.
The exponents of the two bounds converge to $e_{\rm lower}$ and $\kappa_2(\lambda,\sigma)$, respectively, so Lemma~\ref{th:PE_nvR_sandwich} does not determine $P(E|\bm{\theta})$ up to a subexponential factor unless $\theta_1+\theta_2=1$ and $\theta_1-\theta_2=\lambda$ (the equality case of Lemma~\ref{lem:c2_elower}).
The exact exponent $e_{\rm lower}$ is obtained in Proposition~\ref{prop:reliability} from the upper bound \eqref{eq:app_union}, whose exponent converges to $e_{\rm lower}$ by Lemma~\ref{lem:mu_asymptotic}.

\section{Experimental Details}\label{app:exp_details}

\textbf{Column subsets (Figure~\ref{fig:emdr_vs_d}(b)).}
For each dataset the column pool is fixed across seeds: it consists of all $28+56=84$ columns of the 2vR and 3vR codes for CIFAR-10 ($M=8$), all $66+220=286$ columns for CIFAR-100 ($M=12$), and the $120$ columns of the 2vR code together with $200$ randomly chosen columns of the 3vR code ($320$ in total) for ImageNet-1k ($M=16$).
For $K\in\{3,4,5,6,8,10,14,18,24,32,44,60,80,120,180,240,300\}$ (not exceeding the pool size) and up to $8$ random draws per $K$ (tables with repeated code words are discarded), $K$ columns are drawn from the pool, the binary classifiers are trained, and the minimum distance of the drawn table and its sum-decoding error rate are recorded.
The records are binned by the minimum distance with bin edges $0.5,1.5,2.5,3.5,4.5,6.5,8.5,11.5,15.5,20.5,27.5,37.5,50.5,67.5,90.5,120.5$; Figure~\ref{fig:emdr_vs_d}(b) plots the mean distance and the mean of $\Delta_{\rm real}$ in each bin.
The ImageNet-1k images are drawn from a subset of the training set, and the classes are drawn anew in each seed.

\textbf{Decoding rule and reference decoder.}
On CIFAR-100 (ResNet-50 features, PCA to $100$ dimensions, $M\in\{4,6,8,10,12,16\}$, $25$ seeds), the difference between the sum-decoding and LL-decoding error rates of the 2vR code on the same classifiers is statistically equivalent to $0$ for $M=6$--$16$ (TOST equivalence test at significance level $0.05$ with equivalence margin $0.005$, a practically negligible difference in error rate; largest $p=1.2\times10^{-10}$), and exactly $0$ in every seed for $M=4$, where the two decoders make the same decisions.
In the same $150$ conditions, a multi-class logistic regression trained on the same features has a higher error rate than LL decoding of the 2vR classifiers in $136$ conditions, an equal one in $8$ and a lower one in $6$ (mean difference $0.011$).
For $M=4$ the implementation stored each 2vR partition together with its complement ($n=M/2$); since a logistic regression trained on complemented labels outputs $1-p_j$, this duplication does not change the decision of either decoder (up to solver tolerance).

\textbf{3vR versus 2vR.}
The comparison in Section~\ref{subsec:exp_emdr_vs_d} uses the same CIFAR-100 runs.
It is restricted to $M\in\{8,10,12\}$, the values for which $d(3,M)>d(2,M)$: for $M=6$ the 3vR code has $n=M/2$ and $d(3,6)=\binom{4}{2}=6<d(2,6)=8$ by \eqref{eq:d_nvR}, and for $M=4$ a 3vR code lies outside Definition~\ref{def:nvR}.
The seed-matched difference $\hat{P}_{\rm err}^{\rm 3vR}-\hat{P}_{\rm err}^{\rm 2vR}$ under sum decoding has mean $+0.0020$ (Wilcoxon signed-rank test $p=1\times10^{-5}$, $75$ pairs).

\textbf{Effective distance of 2vR and 3vR (Section~\ref{sec:discussion}(iii)).}
A teacher, a 100-class multilayer perceptron on standardized ResNet-50 features reduced by PCA to $256$ dimensions, is trained on $45{,}000$ CIFAR-100 training images, with early stopping and temperature calibration ($T=1.087$) on the remaining $5{,}000$; its test accuracy is $0.709$.
For each seed and the same $M$ classes, images, PCA-100 features and logistic regressions as above, the labels are resampled from the teacher posterior renormalized to the $M$ classes, $\bm{\theta}^*(\bm{x})$, whose Bayes error rate is $0.065$, $0.084$ and $0.103$ for $M=8,10,12$.
Because $\bm{\theta}^*$ is known, we report expected error rates $E[1-\theta^*_{\hat{i}}]$, free of label-sampling noise, and the log-odds residuals $r_j=\ln\frac{p_j}{1-p_j}-\ln\frac{f^*_j(1|\bm{\theta}^*)}{1-f^*_j(1|\bm{\theta}^*)}$.
$\hat\Sigma$ is the covariance of $(r_j)_j$ on $100$ training images per class not used for training, and the predictions are evaluated on the test set by Monte Carlo ($256$ draws per test image).
The seed-matched 3vR$-$2vR differences in excess error are $+0.0003$, $+0.0014$ and $+0.0018$ for $M=8,10,12$ ($25$ seeds each).
The residuals are not independent of the input: regressing the runner-up pair's noise on its margin gives a slope of $-0.29$ to $-0.42$, larger in magnitude for 3vR in every seed, and their excess kurtosis is $0.75$--$1.16$; all predictors treat the residual as independent of the input, which is the likely source of the mismatch in magnitude.
The less confident teacher of Section~\ref{sec:discussion}(iii) doubles the temperature.

\textbf{Scaling in $M$ (Table~\ref{tab:imagenet_pe}).}
We select $M\in\{16,32,64,128,256\}$ ImageNet-1k classes, train the 2vR and logM codes by logistic regression on $200$ training images per class from the ImageNet-1k training set and test on the validation set, using ResNet-50 (pretrained ImageNet1K\_V2) features reduced by PCA to $128$ dimensions ($S=3$ seeds), and report the representative values $M=16,128,256$; $M\leq256$ because the number of 2vR columns becomes impractical beyond $M\approx300$.

\end{document}